\PassOptionsToPackage{dvipsnames}{xcolor}
\documentclass[twocolumn]{aastex631}

\usepackage{physics}
\usepackage{comment}
\usepackage{multirow}

\usepackage{bm}
\usepackage{mathrsfs}

\definecolor{ochre}{rgb}{0.8, 0.47, 0.13}

\begin{document}

\title{Close Encounters of Wide Binaries Induced by the Galactic Tide:\\
Implications for Stellar Mergers and Gravitational-Wave Sources}

\correspondingauthor{Jakob Stegmann}
\email{jstegmann@mpa-garching.mpg.de}

\author[0000-0003-2340-8140]{Jakob Stegmann}
\affiliation{Max-Planck-Institut für Astrophysik, Karl-Schwarzschild-Straße 1, 85741 Garching bei München, Germany}

\author[0000-0003-1817-3586]{Alejandro Vigna-Gómez}
\affiliation{Max-Planck-Institut für Astrophysik, Karl-Schwarzschild-Straße 1, 85741 Garching bei München, Germany}

\author[0000-0001-8789-2571]{Antti Rantala}
\affiliation{Max-Planck-Institut für Astrophysik, Karl-Schwarzschild-Straße 1, 85741 Garching bei München, Germany}

\author[0000-0001-6147-5761]{Tom Wagg}
\affiliation{Department of Astronomy, University of Washington, Seattle, WA 98195, USA}

\author[0000-0003-4818-3400]{Lorenz Zwick}
\affiliation{Niels Bohr International Academy, Niels Bohr Institute, Blegdamsvej 17, 2100 København, Denmark}

\author[0000-0002-6718-9472]{Mathieu Renzo}
\affiliation{Steward Observatory, University of Arizona, 933 N Cherry Avenue, Tucson, AZ 85721, USA}

\author[0000-0001-5484-4987]{Lieke A. C. van Son}
\affiliation{Center for Computational Astrophysics, Flatiron Institute, New York, NY 10010, USA}

\author[0000-0001-9336-2825]{Selma E. de Mink}
\affiliation{Max-Planck-Institut für Astrophysik, Karl-Schwarzschild-Straße 1, 85741 Garching bei München, Germany}
\affiliation{Ludwig-Maximilians-Universität München, Geschwister-Scholl-Platz 1, 80539 München, Germany}

\author[0000-0002-1061-6154]{Simon D. M. White}
\affiliation{Max-Planck-Institut für Astrophysik, Karl-Schwarzschild-Straße 1, 85741 Garching bei München, Germany}



\begin{abstract}
A substantial fraction of stars can be found in wide binaries with projected separations between $\sim10^2$ and~$10^5\,\rm AU$. In the standard lore of binary physics, these would evolve as effectively single stars that remotely orbit one another on stationary Keplerian ellipses. However, embedded in their Galactic environment their low binding energy makes them exceptionally prone to perturbations from the gravitational potential of the Milky Way and encounters with passing stars. Employing a fully relativistic $N$-body integration scheme, we study the impact of these perturbations on the orbital evolution of wide binaries along their trajectory through the Milky Way. Our analysis reveals that the torques exerted by the Galaxy can cause large-amplitude oscillations of the binary eccentricity to $1-e\lesssim10^{-8}$. As a consequence, the wide binary members pass close to each other at periapsis, which, depending on the type of binary, potentially leads to a mass transfer or collision of stars or to an inspiral and subsequent merger of compact remnants due to gravitational-wave radiation. 
Based on a simulation of $10^5$ wide binaries across the Galactic field, we find that this mechanism could significantly contribute to the rate of stellar collisions and binary black hole mergers as inferred from observations of Luminous Red Novae and gravitational-wave events by LIGO/Virgo/Kagra. We conclude that the dynamics of wide binaries, despite their large mean separation, can give rise to extreme interactions between stars and compact remnants.
\end{abstract}


\section{Introduction}\label{sec:Intro}
In recent years, unprecedented astrometric data obtained with the \textit{Gaia} spacecraft have revealed numerous wide stellar binaries with projected separations between $\sim10^2$ and~$10^5\,\rm AU$ \citep{Andrews2017,Oelkers2017,El-Badry2018,Esteban2019,Hartman2020,Tian2020,Hwang2021,El-Badry2021,El-Badry2024}. For instance, \cite{El-Badry2018} identified $\sim5.5\times10^4$ wide main-sequence and white dwarf binaries in the \textit{Gaia} DR2 sample within a distance $d<200\,\rm pc$ to the Sun. Given a local total stellar number density of $n_\star\approx0.1/\,\rm pc^3$ this implies that at least several per cent of all stars in the solar neighborhood and possibly within the Galaxy must be part of a wide binary. Indeed, estimates from the \textit{Gaia} Early Data Release 3 indicate that the wide binary fraction of nearby FGK stars is as large as 10~--~12\,\% \citep{Gaia2021}. 

In the standard lore of binary physics, wide binaries (simply denoted as ``binaries", hereafter) are thought to evolve as non-interacting, effectively single stars \citep[e.g.,][]{Sana2012}. However, this is only true as long as the binaries are assumed to be in isolation. In reality, they are embedded in a galactic environment in which their low binding energy makes them susceptible to perturbations from the gravitational potential of their host galaxy and encounters with passing stars and giant molecular clouds. 
The effect of these perturbations 
has been studied in a wide range of different contexts, e.g., to constrain the existence of a distant companion to our Sun \citep{Antonov1972,Whitmire1984,Davis1984,Hut1984,Weinberg1987,Melott2010,Matese2011,Luhman2014}, to probe the nature of dark matter \citep{Yoo2004,Quinn2009,Monroy2014,Pearrubia2016} and gravity \citep{Pittordis2019,Banik2023}, to explore the stability of binaries with a distant tertiary companion \citep{Kaib2013,Correa-Otto2017,Antonini2017,Michaely2020,Evgeni2022}, and to form colliding stars \citep{Kaib2014} and compact object mergers \citep{Michaely2019,Michaely2020,Raveh2022}.

\begin{figure*}[t]
    \centering
    \includegraphics[width=1\textwidth]{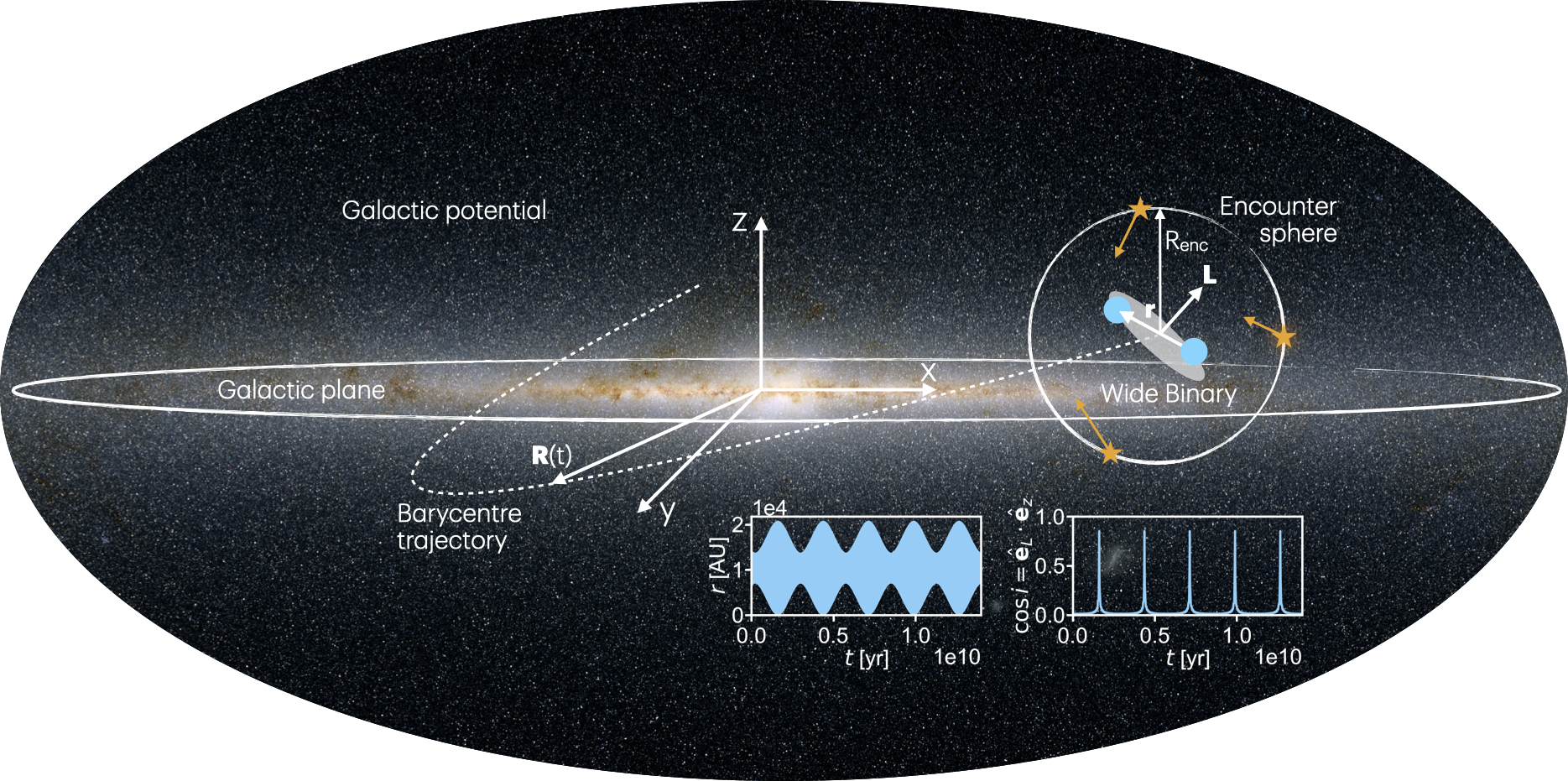}
    \caption{Schematic overview of the problem considered in this work. We study the dynamical evolution of wide binaries along their barycentre trajectory $\mathbf{R}=\mathbf{R}(t)$ through the Galaxy (dashed curve),  denoting its projection onto the Galactic plane as $\bm{\rho}=\bm{\rho}(t)$. We consider one model in which the wide binaries are solely perturbed by the axisymmetric Galactic potential $\Phi$ and another one where we additionally include the impulsive effect of ambient stars impinging on the encounter sphere (solid circle) with radius $R_{\rm enc}$ from the binary barycentre. The main  effect of the Galactic perturbations is to torque the binary orbital angular momentum $\mathbf{L}$ leading to large-amplitude oscillations of the binary eccentricity $e$, and hence its peri- and apoapsis $r=a(1\mp e)$, and the relative orientation $\cos(i)=\hat{\mathbf{e}}_L\cdot\hat{\mathbf{e}}_z$ of the orbital plane w.r.t.~the Galactic frame (insets). The dimensions are not drawn to scale. The Atlas Image of the Milky Way in the near infrared used as a background is courtesy of 2MASS/UMass/IPAC-Caltech/NASA/NSF.}
    \label{fig:sketch}
\end{figure*}

From a dynamical point of view, it is understood that the Galactic perturbations could generally lead to two different extreme outcomes: they may either dissolve the wide binary \citep[e.g.,][]{Weinberg1987,Yan-Fei2010,Correa-Otto2017} or they torque it to an eccentric orbit on which the two binary members pass close to each other at periapsis \citep[e.g.,][]{Heisler1986,Collins2008,Kaib2014,Correa-Otto2017b,Modak2023,Hamilton2023}. These findings were largely made using one or more of the following assumptions: 
\begin{enumerate}
    \item The Galactic potential is averaged over the long orbital period of the binary (also known as secular approximation). \label{item:1}
    \item The gravitational perturbation from the Galactic potential is approximated by the leading-order contribution of an expansion about the binary centre of mass (barycentre). \label{item:2}
    \item Focusing on the dynamics of wide binaries in the solar neighbourhood, the perturbation from the Galactic potential is evolved along the (circular) trajectory of the Sun at a Galactocentric radius $\rho\approx8\,\rm kpc$ from the centre of the Milky Way. \label{item:3}
\end{enumerate}
%
%
However, studies of binaries orbiting in other gravitational potentials let us wonder if these assumptions lead to an underestimation of the perturbative effect of the Galactic potential on wide binaries. For instance, \cite{HamiltonRafikovI,HamiltonRafikovII,HamiltonRafikovIII} and \cite{BubPetrovich} studied the evolution of binaries orbiting inside the smooth potential of a star cluster and found that resonant behaviour can excite their eccentricity to near-unity ($e\rightarrow1$). Similarly, the eccentricity of binaries moving inside the Keplerian potential of a distant tertiary companion (i.e., those that form a hierarchical triple) can undergo a long-term growth to near-unity through the so-called eccentric von Zeipel-Kozai-Lidov effect \citep{Naoz2016}. For either case, it was shown that the external potentials exert a torque on the binaries which could remove nearly all their orbital angular momentum. Depending on the type of binary this could cause a stellar collision \citep{Katz2012,Kushnir2013,Toonen2018,Toonen2020,Stegmann2022b} or a gravitational-wave capture \citep{Antonini2014,Antonini2017,Rodriguez2018} at close periapsis passage. 

In this regard, it would be surprising if the potential of a galaxy acted differently than the aforementioned examples. Indeed, if we consider an example stellar-mass binary (total mass $m=1\,\rm M_\odot$) moving on circular trajectory at a Galactocentric radius $\rho=5\,\rm kpc$ on the Galactic plane and treat the Milky Way as a tertiary point mass equal to the enclosed mass $M(\rho)$ we find that the latter perturbs the binary on a characteristic (von Zeipel-Kozai-Lidov) timescale \citep{Antognini2015}
\begin{align}
    t_{\rm ZKL}=&\,9.0\,{\rm Gyr}\left(\frac{5.9\times10^{10}\,\rm M_\odot}{M(\rho)}\right)\left(\frac{m}{M_\odot}\right)^{1/2}\nonumber\\
    &\times\left(\frac{\rho}{5\,\rm kpc}\right)^{3} \left(\frac{5\times10^3\,\rm AU}{a}\right)^{3/2}. 
\end{align}
While this toy model \citep{Chebotarev1966} indicates that the tertiary ``Galaxy companion" could induce large-amplitude oscillations of the binary eccentricity within a Hubble time for wide semi-major axes $a\gtrsim\mathcal{O}(10^3)\,\rm AU$, it cannot serve as more than a simple back-of-the-envelope estimate to motivate this work. 

Here, we develop a numerical method to evolve the relativistic dynamics of binaries following the mutual gravitational interaction between the two members, the force exerted by the Galaxy in which they orbit, and the effect of encountering stars. Crucially, to accurately evolve binaries which are torqued to a very eccentric orbit we employ a direct $N$-body approach rather than any secular code that derives from a time- (e.g., orbit-)~averaged Hamiltonian. While the secular approximation is oftentimes used to efficiently evolve a large number of, e.g., hierarchical triples \citep[e.g.,][]{Naoz2013,Rodriguez2018,Liu2019,Stegmann2022} and binaries inside clusters \citep[e.g.,][]{Hamilton2019,BubPetrovich} it has been shown that it tends to underestimate the maximum possible eccentricity because it washes out any short-term variability of the torque exerted by the perturbing potential \citep{Antonini2014,Antonini2017,Hamilton2023b}. These shortcomings are circumvented by using a full direct $N$-body integration scheme which, in our case, adds little computational costs because of the low number of particles ($N=2$).

The aim of this work is to use this integrator which relies on none of the aforementioned assumptions (\ref{item:1}~--~\ref{item:3}) in order to systematically and accurately study the dynamical evolution of wide binaries throughout the Galaxy. The primary focus is to understand their dynamical evolution treating the binary members as point masses. We highlight potential implications for wide binary \textit{stars} and wide binary \textit{compact objects}, where stellar tides or natal kicks also may play a role, but defer a thorough investigation to a follow-up study.

This paper is organised as follows. In Sec.~\ref{sec:Methods}, we present our method to evolve the wide binaries. 
In Sec.~\ref{sec:initial}, we describe how we initiate a population of wide binaries whose dynamical evolution we simulate along their trajectory through the host galaxy. In Sec.~\ref{sec:results}, we present the results of their numerical integration. Finally, our findings are summarised and discussed in Sec.~\ref{sec:discussion}. 

\section{Methods}\label{sec:Methods}
\subsection{Equations of motion and numerical integration}\label{sec:eom}
We study the dynamical evolution of wide binaries as they move through the smooth gravitational potential $\Phi$ of their host Galaxy. Denoting the masses, positions, and velocities of the two binary members as $m_i$, $\mathbf{r}_i$, and $\mathbf{v}_i$ ($i=1,2$), respectively, the equations of motion can be written as\footnote{Throughout this work, the magnitude, unit vector, and time derivative of some vector $\bm{V}$ are written as $V=\left\|\bm{V}\right\|$, $\hat{\bm{e}}_V=\bm{V}/V$, and $\bm{\dot{V}}=d\bm{V}/d t$, respectively. $G$ refers to the gravitational constant.}
\begin{align}
    \frac{d \mathbf{r}_i}{d t}&=\mathbf{v}_i,\label{eq:eom1}\\
    \frac{d \mathbf{v}_i}{d t}&=\mathbf{a}_i+\mathbf{g}_i+\mathbf{f}_i\label{eq:eom2},
\end{align}
where
\begin{equation}
    \mathbf{a}_i=Gm_j\frac{\mathbf{r}_j-\mathbf{r}_i}{\|\mathbf{r}_j-\mathbf{r}_i\|^3}\quad(i\neq j)
\end{equation}
is the Newtonian acceleration due to the binary companion and $\mathbf{f}_i=-\nabla\Phi|_{\mathbf{r}_i}$ is the acceleration due to the external Galactic potential. Relativistic corrections to the internal binary motion are captured by the velocity-dependent post-Newtonian accelerations $\mathbf{g}_i = \mathbf{g}_i(\mathbf{r}_j-\mathbf{r}_i,\mathbf{v}_j-\mathbf{v}_i)$ with $i\neq j$.  We can characterise the orbit of the binary in terms of its angular momentum vector $\mathbf{L}$ and eccentricity vector $\mathbf{e}$ which are
\begin{align}
    \mathbf{L}&=\mu\mathbf{r}\times\mathbf{v},\\
    \mathbf{e}&=\frac{\mathbf{v}\times(\mathbf{r}\times\mathbf{v})}{Gm}-\frac{\mathbf{r}}{r},
\end{align}
where $\mathbf{r}=\mathbf{r}_2-\mathbf{r}_1$ and $\mathbf{v}=\mathbf{v}_2-\mathbf{v}_1$ are the relative separation and velocity vectors, and $m=m_1+m_2$ and $\mu=m_1m_2/m$ are the total and reduced mass of the binary, respectively. They have magnitudes $L=\mu\sqrt{Gma(1-e^2)}$ and $\|\mathbf{e}\|=e$, where $a$ and $e$ are the semi-major axis and eccentricity of the orbit, respectively. Erecting a frame in which the Galaxy is centred at $x=y=z=0$ and defining the Galactic plane as $z=0$, we denote the position of the binary barycentre as $\mathbf{R}=(R_x,R_y,R_z)^{\rm T}$ and its projection onto the Galactic plane as $\bm{\rho}=(R_x,R_y,0)^{\rm T}$. Fig.~\ref{fig:sketch} shows a schematic overview of our considered geometry.  

In the absence of any perturbative force ($\mathbf{f}_i=\mathbf{g}_i=0$) the equations of motion~\eqref{eq:eom1}~--~\eqref{eq:eom2} reduce to that of a Keplerian orbit which preserves $\mathbf{L}$ and $\mathbf{e}$. As we will see, however, the forces exerted by the Galactic potential and stellar encounters (Sec.~\ref{sec:encounters}) can put the binary members on a nearly radial orbit ($e\rightarrow1$) on which they pass each other at an extremely close pericentre distance $r=a(1-e)\ll a$ despite their wide semi-major axis. To accurately recover their evolution it is necessary that we also include post-Newtonian corrections $\mathbf{g}_i$ to Eq.~\eqref{eq:eom2} which we do to 3.5 order \citep{MoraWill2004}. 

To evolve the equations of motion~\eqref{eq:eom1} and~\eqref{eq:eom2} we use the publicly available direct $N$-body integrator {\tt MSTAR} \citep{Rantala2020,Mannerkoski2023} which is based on an algorithmically regularized integration technique \citep[e.g.,][]{Mikkola1999,Preto1999,Mikkola2008,Hellstrom2010,Trani2023}. The time-transformed versions of the equations of motion~\eqref{eq:eom1} and~\eqref{eq:eom2} that {\tt MSTAR} integrates are Eqs.~(7) and~(8) in \cite{Rantala2020}. The Gragg-Bulirsch-Stoer (GBS) extrapolation technique \citep{Gragg1965,Bulirsch1966} is used to achieve an extremely high integration accuracy. Briefly, in the GBS method a longer timestep $\Delta t$ is subdivided into $n$ substeps with lengths $\Delta t / n$ and integrated with a suitable numerical integrator, in our case the leapfrog. As $n$ is increased, the results will in general converge towards the exact solution of the equations of motion over $\Delta t$. Finally, the results are extrapolated to $n \rightarrow \infty$ using a polynomial or rational function extrapolation. We use a Gragg-Bulirsch-Stoer error tolerance parameter of $\eta_\mathrm{GBS}=10^{-12}$. The {\tt MSTAR} endtime iteration tolerance parameter is set to $\eta_t = 10^{-6}$. For more details of {\tt MSTAR} and the algorithmic regularization see \cite{Rantala2020} and the appendices of \cite{Rantala2017}. In particular, by accounting for the post-Newtonian corrections, {\tt MSTAR} has the capability to follow the inspiral and the subsequent merger of the binary members due to the emission of gravitational waves until they come as close as twelve combined Schwarzschild radii.

\subsection{Gravitational potential of the Galaxy}\label{sec:potential}
We have modified {\tt MSTAR} to also include the ability to follow the position of the binary barycentre and 
the relative acceleration due to the external Galactic potential as in Eqs.~\eqref{eq:eom1} and~\eqref{eq:eom2}. Here, we adopt an axisymmetric mass-model for the Milky Way from \citet{gala} consisting of a spherical nucleus and bulge, a spherical NFW dark matter halo, and an exponential disk. The numerical values for the parameters of all components of the potential which are explicated in the following are adopted from the model of \citet{Price-Whelan+2024:2022zndo....593786P} which is based on the kinematic Milky Way analysis of \citet{Eilers2019} and \citet{Darragh-Ford2023}. Specifically, the potentials of the nucleus and the bulge as a function of the binary member positions $\mathbf{r}_i$ are modelled with a \cite{Hernquist1990} profile 
\begin{equation}
    \Phi_{\rm Hernquist}(\mathbf{r}_i)=-\frac{GM}{r_i+c},\label{eq:Hernquist}
\end{equation}
where for the nucleus we use $M=1.814\times10^9\,\rm M_\odot$ and $c=6.889\times10^{-2}\,\rm kpc$ and for the bulge $M=5.0\times10^9\,\rm M_\odot$ and $c=1.0\,\rm kpc$. The potential of the NFW dark matter halo is given by \citep{NFW1996}
\begin{equation}
    \Phi_{\rm NFW}(\mathbf{r}_i)=-\frac{GM}{r_s}\frac{\ln(1+r_i/r_s)}{r_i/r_s},
\end{equation}
where $M=5.542\times10^{11}\,\rm M_\odot$ and $r_s=15.626\,\rm kpc$. The stellar disk is constructed from a sum of three \cite{Miyamoto1975} disk potentials which models the thin and thick disk component of the Milky Way \citep{Smith2015}
\begin{equation}
    \Phi_{\rm Miyamoto}(\mathbf{r}_i)=-\frac{GM}{\sqrt{x_i^2+y_i^2+(a+\sqrt{z_i^2+b^2})^2}},\label{eq:Miyamoto}
\end{equation}
where $x_i$, $y_i$, and $z_i$ are the Cartesian components of the binary member positions, $M=7.872\times10^9\,\rm M_\odot$, $a=1.526\,\rm kpc$, and $b=0.207\,\rm kpc$ for the first instance of the potential, $M=-2.756\times10^{11}\,\rm M_\odot$, $a=6.783\,\rm kpc$, and $b=0.207\,\rm kpc$ for the second, and $M=3.206\times10^{11}\,\rm M_\odot$, $a=5.895\,\rm kpc$, and $b=0.207\,\rm kpc$ for the third. 

The total potential of our axisymmetric Milky Way model is then given by the sum of all components described by Eqs.~\eqref{eq:Hernquist}~--~\eqref{eq:Miyamoto}. For the purpose of this work, we are ignoring more complex non-axisymmetric features of our Galaxy such as its bar, spiral arms, and molecular cloud structure and neglecting any time-evolution of the potential. We discuss potential implications of these simplifications in Sec.~\ref{sec:discussion}.

\subsection{Stellar encounters}\label{sec:encounters}
In addition to the force caused by the smooth potential of the Galaxy, a wide binary experiences gravitational perturbations from encounters with passing stars. Assuming that the binary moves through a locally homogeneous and isotropic sea of stellar perturbers the rate of encounters within a distance $R_{\rm enc}$ to its barycentre can be written as \citep{Hamers2021}
\begin{align}\label{eq:rate-encounters}
    \Gamma_{\rm enc}&\approx2\sqrt{2\pi}R_{\rm enc}^2\sigma_\star n_\star\nonumber\\
    &\approx\frac{60.2}{\rm Gyr}\left(\frac{R_{\rm enc}}{10^4\,\rm AU}\right)^2\left(\frac{\sigma_\star}{50\,\rm km\,s^{-1}}\right)\left(\frac{n_\star}{0.1/\rm pc^3}\right),
\end{align}
where $\sigma_\star$ and $n_\star$ are the relative velocity dispersion and number density of the perturbers, respectively \citep{Flynn2006}. Eq.~\eqref{eq:rate-encounters} shows that the rate of encounters quickly grows if we consider distances $R_{\rm enc}$ much larger than the typical relative separation of the wide binaries and modelling each of those would become computationally expensive.
Fortunately, the effect on wide binaries is by far dominated by encounters which are ``penetrative", ``weak", and ``impulsive" \citep{Hamilton2023}. This means that the perturbation from the few stars which get as close as $\sim a$ to the binary (so-called ``penetrative" encounters) is much stronger than the cumulative effect of the many more distant encounters, most encounters are harmless, meaning that they unlikely disrupt the binary (``weak"), and that the stellar perturbers encounter the binary on a timescale much shorter than its orbital period $T=2\pi\sqrt{a^3/Gm}$ (``impulsive") which is due to the fact that the typical velocity dispersion $\sigma_\star$ of ambient stars is much larger than the orbital velocity of the binary
\begin{align}\label{eq:orbital-velocity}
v_{{\rm orb}} = 0.9\,{\rm km}\,{\rm s}^{-1}\, \left (\frac{m}{\rm M_\odot} \right )^{1/2} \left ( \frac{a}{10^3\,{\rm AU}} \right )^{-1/2}.
\end{align}
This allows us to ignore any encounter at a distance $R_{\rm enc}\gg a$ and to only consider close encounters which in the impulsive approximation impart an instantaneous velocity kick to the two stars of the binary. 

The effect of those encounters is implemented as following. After each successful internal integration step $\Delta t$ of {\tt MSTAR} we sample a number of encounters within $\Delta t$. We do so by drawing a random number $k$ of encounters from a Poisson distribution with mean $\lambda=\Gamma_{\rm enc}\Delta t$. The number density $n_\star$ that determines $\Gamma_{\rm enc}$ is calculated from the current position of the binary within the Galactic disk, assuming a single perturber mass $m_p=1\,\rm M_\odot$. Furthermore, we use $\sigma_\star=50\,\rm km\,s^{-1}$ and only consider encounters within $R_{\rm enc}=100r$ where $r$ is the current relative separation of the binary. 

Each of the $k$ encounters imparts an instantaneous velocity kick to both members of the binary which is computed following \citet{HamersTremaine2017} and \citet{Hamers2021} as summarised below.

\begin{enumerate}
    \item In an inertial frame centred at the binary barycentre, we sample a random perturber position $\mathbf{R}_{\rm enc}$ at a distance $R_{\rm enc}$ from the binary barycentre. We give the stellar perturber a random relative velocity $\mathbf{V}_{\rm enc}$ impinging on the encounter sphere, which is drawn from a a distribution $P(\mathbf{V}_{\rm enc})\propto\exp(-V_p^2/2\sigma_\star^2)H(-\mathbf{V}_{\rm enc}\cdot\hat{\mathbf{R}}_{\rm enc})(-\mathbf{V}_{\rm enc}\cdot\hat{\mathbf{R}}_{\rm enc})$, where $H$ is the Heaviside step function \citep{Henon1972}.
    \item We calculate the impact parameters which describe how close the perturber gets to the binary barycentre and either of the binary members ($i=1,2$), respectively, as
    \begin{align}
    \mathbf{b}&=\mathbf{R}_{\rm enc}-\hat{\mathbf{V}}_{\rm enc}\left({\mathbf{R}}_{\rm enc}\cdot\hat{\mathbf{R}}_{\rm enc}\right),\\
    \mathbf{b}_{i}&=\mathbf{b}-\mathbf{R}_{i}-\hat{\mathbf{V}}_{\rm enc}\left[\left(\mathbf{b}-\mathbf{R}_{i}\right)\cdot\hat{\mathbf{V}}_{\rm enc}\right],
    \end{align}
    where $\mathbf{R}_{i}$ are the positions of the binary members in the barycentre frame.
    \item We add an instantaneous kick 
    \begin{equation}
        \Delta v_i=\frac{2Gm_p}{V_{\rm enc}}\frac{\hat{\mathbf{b}}_i}{b_i}
    \end{equation}
    to the velocity of both binary members and update the barycentre velocity accordingly. In the impulsive approximation the instantaneous positions of the binary members and their barycentre remain unaffected.
\end{enumerate}
We have modified {\tt MSTAR} to include the kick prescription above. Typically, the integration step $\Delta t$ of {\tt MSTAR} is a small fraction of the orbital period of the binary and within this step it experiences $k\sim\mathcal{O}(1$~--~$1000$) weak encounters.

\begin{figure*}[t]
    \centering
    \includegraphics[width=.9\textwidth]{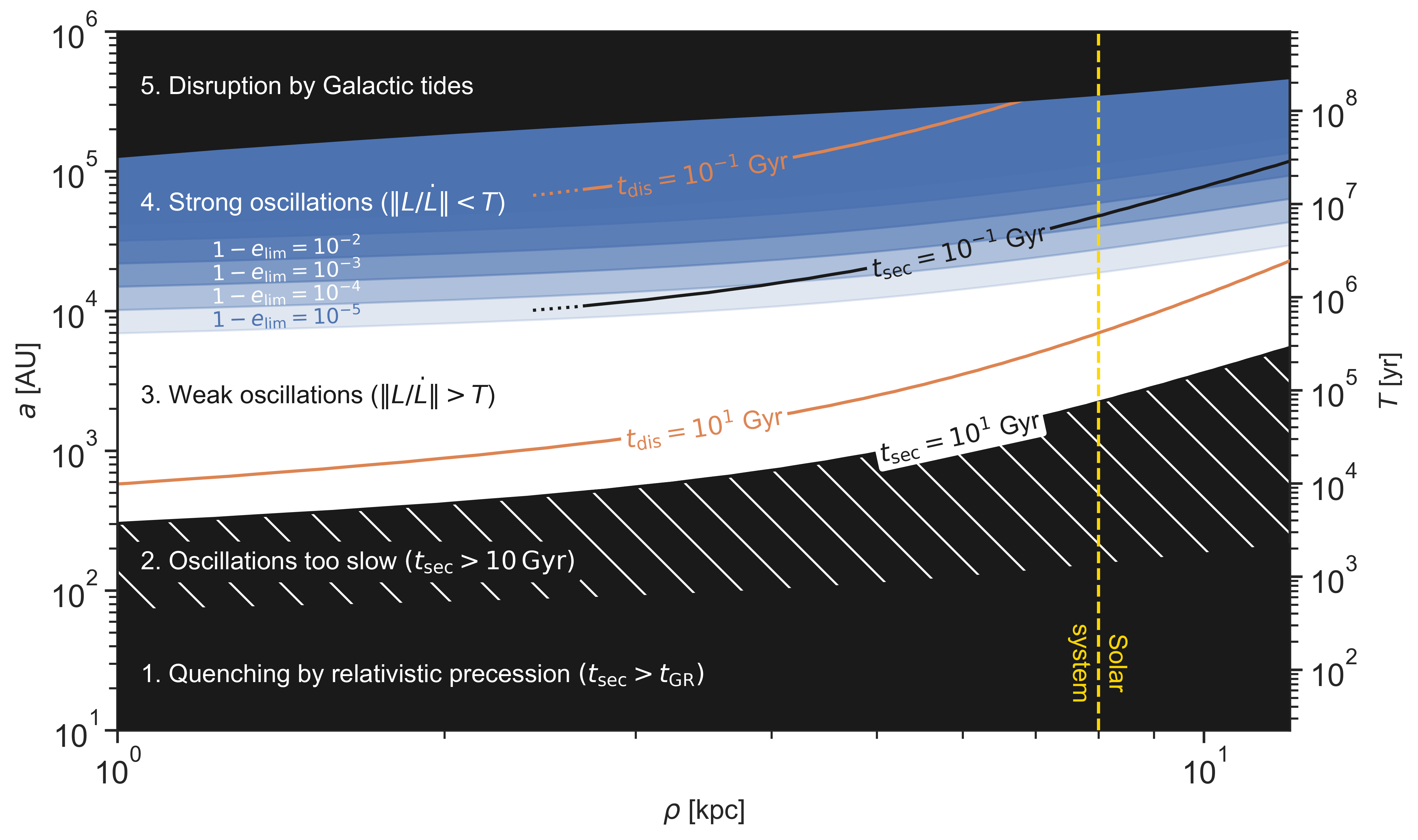}
    \caption{Five different dynamical regimes for wide binaries inside the Milky Way potential. We assume an example binary with semi-major axis $a$ whose barycentre moves on a circular trajectory at a Galactocentric radius $\rho$ on the Galactic plane. We set the binary masses to $m_1=m_2=1\,\rm M_\odot$ but note that the displayed dynamical regimes are not particularly sensitive to this choice. The figure shows that our sampled range, $a=10^2$~--~$10^5\,\rm AU$ (see Sec.~\ref{sec:initial}), 
    fully covers the spectrum where the Galactic potential can influence wide binaries on a meaningful timescale. The details of the figure are fully explained in Sec.~\ref{sec:results}.}
    \label{fig:timescales}
\end{figure*}

\subsection{Initial conditions}\label{sec:initial}
Using the integration scheme described above, we simulate the evolution of a population of $10^5$ wide binaries whose initial parameters (denoted with subscript ``0") are as follows. The semi-major axes $a_0$ are drawn from a log-uniform distribution between $10^2$ and $10^5\,\rm AU$ and the eccentricities from a thermal distribution $p(e_0)\propto e_0$ between $0$ and $1$. Our choice of the semi-major axis range roughly corresponds to the regime where secular changes of the binaries could be induced within a Hubble time, see Sec.~\ref{sec:results} for details. While in reality the mass distributions of wide low-mass and high-mass stellar binaries and binaries of compact objects may significantly differ from each other \citep{Moe2017,El-Badry2024} we opt to choose a mass function which is as simple as possible in order to unbiasedly focus on its consequences for the dynamics. Thus, the component masses $m_{1,2}$ are independently sampled from a uniform distribution between $1$ and $10\,\rm M_\odot$ which covers the typical masses of solar- to B-type stars and the peak of the mass distribution of binary black hole mergers inferred from gravitational-wave detections \citep{Abbott2023}. We discuss in Sec.~\ref{sec:discussion} how plausible it is to form wide binary black holes from wide massive binary stars. 

The orientation of the binary orbits with respect to the Galaxy are sampled isotropically, i.e., the initial argument of periapsis $\omega_0$, longitude of the ascending node $\Omega_0$, cosine of the inclination $\cos i_0$, and orbital phase $f_0$ are sampled randomly and define the initial positions and velocities of the binaries as \citep{Merritt2013}
\begin{align}
    \mathbf{r}_0=&r_0[\mathbf{u}_1\cos(f_0+\omega_0)+\mathbf{u}_2\sin(f_0+\omega_0)],\\
    \dot{\mathbf{r}}_0=&\sqrt{\frac{Gm}{p_0}}\{-\mathbf{u}_1[e_0\sin\omega_0+\sin(f_0+\omega_0)]\nonumber\\
    &+\mathbf{u}_2[e_0\cos\omega_0+\cos(f_0+\omega_0)]\},
\end{align}
where $p_0=a_0(1-e_0^2)$ and $r_0=p_0/(1+e_0\cos f_0)$ and
\begin{align}
    \mathbf{u}_1=\begin{pmatrix}
    \cos\Omega_0\\
    \sin\Omega_0\\
    0
    \end{pmatrix},\quad
    \mathbf{u}_2=\begin{pmatrix}
    -\cos i_0\sin\Omega_0\\
    \cos i_0\cos\Omega_0\\
    \sin i_0
    \end{pmatrix}.
\end{align}
The Galactocentric positions and velocities of the binary barycentres are initialised using \texttt{cogsworth}\footnote{\url{https://cogsworth.readthedocs.io/}} (Wagg et al.~in prep.) which is based on the Galaxy model of \citet{Wagg2022}. This is an empirically informed model of the metallicity-dependent star formation history of the Milky Way in a low-$\alpha$ and high-$\alpha$ disc and bulge component \citep{McMillan+2011:2011MNRAS.414.2446M, Bovy+2016:2016ApJ...823...30B, Bovy+2019:2019MNRAS.490.4740B,Frankel+2018:2018ApJ...865...96F}. We initialise the trajectory of the binary barycentre within the Galaxy based on its initial position, such that the initial velocity is equal to its circular velocity in the Milky Way potential with an additional isotropic $5 \, {\rm km\,s^{-1}}$ dispersion \citep{gala, Price-Whelan+2024:2022zndo....593786P}. 

The evolution of the binaries which are initiated in this way is simulated twice. In one simulation we focus exclusively on the influence from Galactic tides (Sec.~\ref{sec:potential}) and ignore the effect of stellar encounters (Sec.~\ref{sec:encounters}). In the other simulation we repeat the integration of each system but consider the combined effect of Galactic tides and stellar encounters. In either case, the binaries are evolved for a maximum integration time $t_{\rm max}=14\,\rm Gyr$ unless they merge earlier due to the emission of gravitational waves or are disrupted in the sense that their semi-major axis becomes larger than the initial value by a factor of ten. Binaries that would undergo a collision or mass transfer if they were made of stars are identified by post processing the data as described below in Sec.~\ref{sec:results}.

\section{Results}\label{sec:results}
\begin{figure*}[t]
    \centering
    \includegraphics[width=1\textwidth]{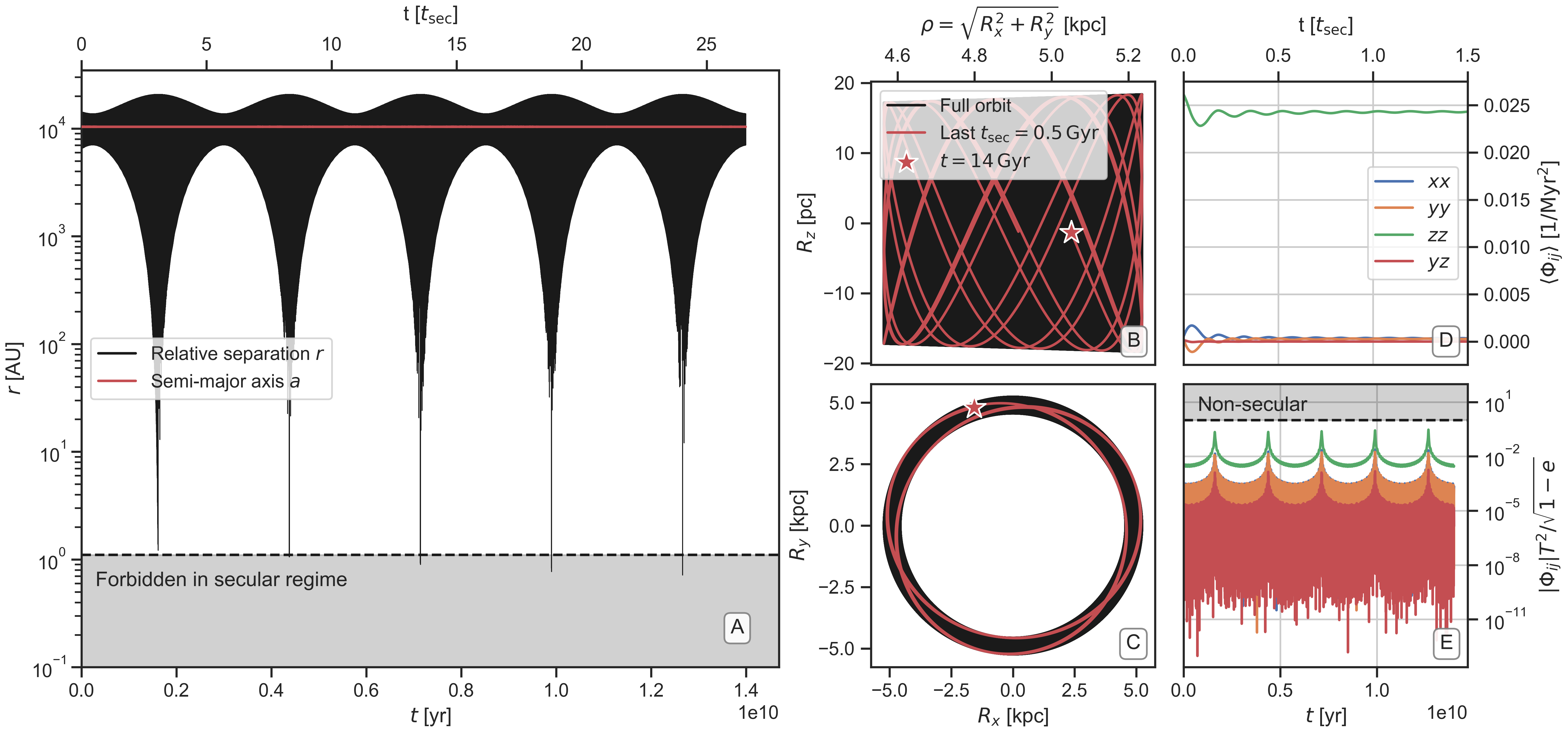}
    \caption{Example of a wide binary orbiting inside the Milky Way potential without stellar encounters. {panel~A}: Relative separation $r$ of the binary members and semi-major axis $a$ as a function of time. The grey-shaded area is defined by $r<a(1-e_{\rm max,sec})$ where $e_{\rm max,sec}$ is given by Eq.~\eqref{eq:emax-sec}. For the secondary abscissa we calculate the secular timescale $t_{\rm sec}$ from Eq.~\eqref{eq:tsec}. {Panels~B} and~C: The trajectory of the binary barycentre viewed edge-on and face-on, respectively. The black curve shows the entire trajectory over $t=14\,\rm Gyr$, red highlights the last $t\in[14\,\rm Gyr-t_{\rm sec},14\,\rm Gyr)$ of it, and the star symbol indicates the final ($t=14\,\rm Gyr$) position of the binary. {panel~D}: The time-average $\langle\Phi_{ij}\rangle(t)$ of the tidal field components along the barycentre trajectory. Off-diagonal terms not shown here are zero or quickly decay to it. {Panel~E}: The variations of the tidal field to check for the secular break-down condition in Eq.~\eqref{eq:secular-condition}. The grey-shaded area defines the parameter space where the condition is satisfied and large non-secular eccentricities are possible.
    } 
    \label{fig:secular-torus-filling}
\end{figure*}
\begin{figure*}[t]
    \centering
    \includegraphics[width=1\textwidth]{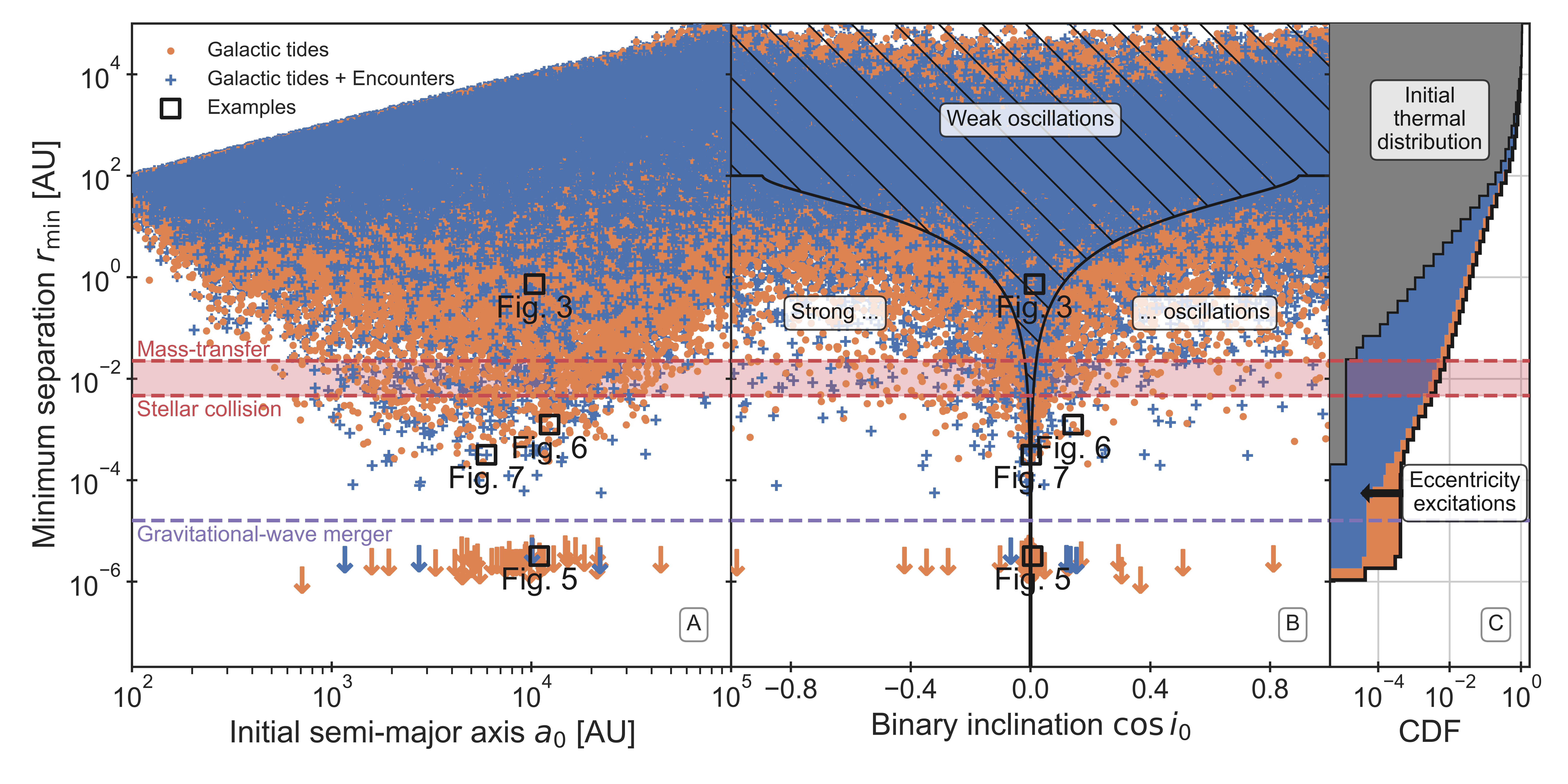}
    \caption{The effect of Galactic tides in driving wide binaries to close separations. {Panels~A} and~B show the minimum separation $r_{\rm min}$ of the binaries in our population as a function of their initial semi-major axis $a_0$ and the initial inclination $\cos i_0$ between their orbital plane and the Galactic plane, respectively. The minimum separation $r_{\rm min}$ is defined as the closest distance the binary members ever achieve due to Galactic tides over a maximum integration time of $14\,\rm Gyr$, either when considering (orange dots) or when neglecting (blue crosses) encounters with passing stars. The purple dashed line separates compact object binaries that merge due to the emission of gravitational waves, where arrows indicate that the binaries are simulated until merging. The red-shaded area shows the regime in which stellar binaries fill their Roche-lobes at periapsis \citep{Eggleton1983}; below that area they potentially collide. The black line in {panel~B} forms a narrow cone defining the smallest minimum separation allowed in the torus-averaged approximation (Eq.~\eqref{eq:emax-sec}) for a fiducial semi-major axis $a_0=10^2\,\rm AU$. Black squares indicate particular example systems shown in Figs.~\ref{fig:secular-torus-filling} and~\ref{fig:non-secular-torus-filling}~--~\ref{fig:secular-encounters}. {Panel~C} shows the cumulative distribution functions (CDF) as well as the initial distribution of $r_{\rm min}=a_0(1-e_0)$ following a thermal eccentricity distribution (see Sec.~\ref{sec:initial}).
    }
    \label{fig:rmin-r0}
\end{figure*}
A wide binary and its host galaxy can be viewed as an effective hierarchical triple in which the ``inner" orbit of the binary members around each other and the much larger ``outer" orbit of the binary barycentre in the Galaxy exchange orbital angular momentum \citep[e.g.,][]{Naoz2016}. Similar to actual triples the tidal torque exerted along the ``outer" orbit inside the Galactic potential can induce incremental changes of the binary angular momentum $L\propto\sqrt{1-e^2}$ per ``inner" orbital period. Over longer timescales these changes can accumulate and give rise to large-amplitude oscillations of the binary eccentricity $e$ which determines how close and far the two binary members get at peri- and apoapsis $r=a(1\mp e)$, respectively. This dynamical mechanism is studied in detail in the following subsections. In order to facilitate this investigation, we outline in Fig.~\ref{fig:timescales} the most relevant dynamical regimes for wide binary evolution which are as follows: 

\begin{enumerate}
    \item \textit{Quenching by relativistic precession:} In close binaries located in the lower black-shaded area ($a\lesssim\mathcal{O}(10^2)\,\rm AU$), the effect of the Galactic tides is suppressed by the ``fast" relativistic precession of the binary orbit \citep{Schwarzschild1916}. Formally, this is the case when the so-called secular timescale $t_{\rm sec}$ at which the Galaxy perturbs the binary (which we define in Sec.~\ref{sec:torus-secular}) exceeds the precession timescale $t_{\rm GR}=a^{5/2}c^2/3G^{3/2}m^{3/2}$ \citep[e.g.,][]{Liu2015}.
    \item \textit{Oscillations too slow:} If the semi-major axis is a few $10^2\,\rm AU$ (hatched area), the Galaxy-induced eccentricity oscillations are no longer quenched by relativistic effects but occur on a long timescale $t_{\rm sec}>10^1\,\rm Gyr$ which renders them inefficient at perturbing the binary within a Hubble time.
    \item \textit{Weak oscillations:}\label{regime:weak} In the white-shaded region, the torques do act on a timescale shorter than the Hubble time but it is much longer than the orbital period ($\|L/\dot{L}\|>T$). In this regime, they can incrementally perturb the binary orbit-by-orbit and slowly remove (or return) the binary orbital angular momentum (secular approximation). As a result, the Galactic torques give rise to regular long-term eccentricity oscillations whose amplitudes are relatively mild (typically $1-e\gtrsim10^{-4}$), unless the binary orbital plane is near-perpendicular to the Galactic plane in which case extreme eccentricities are possible.
    \item \textit{Strong oscillations:}\label{regime:strong} In the blue-shaded regime, the torques act on a timescale shorter than the orbital period ($\|L/\dot{L}\|<T$). This corresponds to a ``secular break-down" which renders the secular approximation formally inapplicable. As a result, we find that extreme eccentricity excitations are possible regardless of the binary inclination. The exact boundary of this regime depends on the current eccentricity of the binaries. Those which are already eccentric can be further driven towards extreme eccentricities more easily, which is indicated by the light-blue shading ($e=e_{\rm lim}$).
    \item\textit{Disruption by Galactic tides:} Ultra-wide binaries located in the upper black-shaded region ($a\gtrsim\mathcal{O}(10^5)\,\rm AU$) tend to be ripped apart quickly by the Galactic tides (Hill instability) and are therefore short-lived \citep{Yan-Fei2010}.
\end{enumerate}

In summary, the parameter space of interest where Galactic torques can induce significant eccentricity oscillations of binaries on a meaningful timescale is defined by the regimes~\ref{regime:weak} and~\ref{regime:strong} which are separately investigated in the following two subsections~\ref{sec:torus-secular} and~\ref{sec:non-secular}. In the entire parameter space the timescale $t_{\rm dis}$ (orange solid lines) for disruption due to the cumulative effect of encountering stars \citep{BinneyTremaine} is longer than $t_{\rm sec}$ which indicates that Galactic tides have typically enough time to significantly perturb the binary orbit before a disruption due to stellar encounters could take place. We study the effect of the latter in the subsections~\ref{sec:results-encounters} and~\ref{sec:disruptions}.

\subsection{Weak oscillations}\label{sec:torus-secular}

Panel~A of Fig.~\ref{fig:secular-torus-filling} shows an example of a wide binary whose binary separation oscillates between $r\approx2\times10^4$ and $1\,\rm AU$ while its barycentre moves on a trajectory shown in {panel~B} (edge-on) and {panel~C} (face-on) at a Galactocentric radius of ${\rho}=\sqrt{R_x^2+R_y^2}\approx5\,\rm kpc$ near the Galactic plane ($R_z\lesssim20\,\rm pc$). Additionally, panels~D and~E display information about the components of the tidal field $\Phi_{ij}=\partial^2\Phi/\partial x_i\partial x_j$ (where $x_1=x$, $x_2=y$, and $x_3=z$) along the barycentre trajectory which determine the leading-order contribution to the Galactic acceleration $\mathbf{f}_i$ in Eq.~\eqref{eq:eom2}. Panel~D shows how the time-average 
\begin{equation}
    \langle\Phi_{ij}\rangle(t)=\frac{1}{t}\int_0^t\Phi_{ij}(\mathbf{R}(t'))\,dt'
\end{equation}
evolves along the trajectory $\mathbf{R}=\mathbf{R}(t)$ for $t>0$. In {panel~E} we show the instantaneous components $\Phi_{ij}(t)=\Phi_{ij}(\mathbf{R}(t))$ (in units of $\sqrt{1-e}/T^2$) along the trajectory.

For this system, the angular frequencies $\dot{\Omega}_\rho$, $\dot{\Omega}_\phi$ and $\dot{\Omega}_z$ of the barycentre trajectory along the radial, azimuthal, and vertical direction, respectively, are much higher than the frequency $1/t_{\rm sec}$ (defined below) of secular changes of the orbital elements and they are non-commensurable, i.e., the binary traces a non-repeating path through the Galaxy \citep{BinneyTremaine}. 
As a consequence, the trajectory of the binary barycentre densely fills an axisymmetric torus within a few $t_{\rm sec}$ and the leading-order perturbation of the binary can be approximately determined by the torus-average of the tidal field $\langle\Phi_{ij}\rangle$ over many azimuthal periods. This situation was generally studied by \cite{HamiltonRafikovI,HamiltonRafikovII,HamiltonRafikovIII} who showed that the torus-averaged tidal field is diagonal ($\langle\Phi_{ij}\rangle=0$ for $i\neq j$ and $\langle\Phi_{xx}\rangle=\langle\Phi_{yy}\rangle$; see {panel~D}) and admits an integral of motion $\Theta=(1-e^2)\cos^2i$. Thus, the eccentricity and the binary orientation oscillate at the expense of each other, with the eccentricity reaching a maximum given by 
\begin{equation}
    1-e_{\rm max,sec}^2=\frac{\Sigma+\sqrt{\Sigma^2-10\Gamma\Theta(1+5\Gamma)}}{1+5\Gamma},\label{eq:emax-sec}
\end{equation}
on a characteristic timescale 
\begin{equation}\label{eq:tsec}
    t_{\rm sec}=\frac{4\pi}{3T\langle\Phi_{zz}+\Phi_{xx}\rangle},
\end{equation}
where
\begin{align}
    \Gamma&=\frac{1}{3}\frac{\langle\Phi_{zz}\rangle+\langle\Phi_{xx}\rangle}{\langle\Phi_{zz}\rangle-\langle\Phi_{xx}\rangle},\\
    \Sigma&=\frac{1+5\Gamma}{2}+5\Gamma\Theta+\left(\frac{5\Gamma-1}{2}\right)D,\\
    D&=e^2\left(1+\frac{10\Gamma}{1-5\Gamma}\sin^2i\sin^2\omega\right).
\end{align}
For our barycentre trajectories we typically find that $\Gamma\approx1/3$ which agrees with the analytical result for epicyclic orbits in disk-like potentials \citep{Hamilton2019}. Thus, in the torus-filling approximation the maximum of the eccentricity and the timescale of oscillations can be calculated from the initial values of $e,i$, and $\omega$ and the average $\langle\Phi_{ii}\rangle$ along the barycentre trajectory. We numerically verify that the example system shown in Fig.~\ref{fig:secular-torus-filling} satisfies this approximation as the timescale for oscillations agrees with $t_{\rm sec}$ within a factor of order unity and the minimum separation is limited by $r\gtrsim a(1-e_{\rm max,sec})$ (indicated by the grey-shaded area in {panel~A}).

In general, reaching extreme eccentricities in the torus-filling approximation is confined to a narrow window around $\cos^2i_0\approx0$, i.e., binaries whose orbital plane is near-perpendicular to the Galactic plane. Indeed, considering initially circular binaries ($e_0=0$) Eq.~\eqref{eq:emax-sec} simplifies to
\begin{equation}
    e_{\rm max,sec}=\sqrt{1-\frac{10\Gamma}{1+5\Gamma}\cos^2i_0},
\end{equation}
which, e.g., for a typical value of $\Gamma\approx1/3$ requires $\|\cos i_0\|\lesssim1.6\times10^{-4}$ to reach an eccentricity as large as $e_{\rm max,sec}\gtrsim1-10^{-4}$ like the one presented in Fig.~\ref{fig:secular-torus-filling}.

In Fig.~\ref{fig:rmin-r0}, we show the minimum separations of our entire binary population, as a function of their initial semi-major axis and the initial mutual inclination with respect to the Galaxy. Here, we focus on the model where we include only the effect of Galactic tides (blue) and investigate the additional effect of stellar encounters (orange) in Sec.~\ref{sec:results-encounters}. Considering the dependence of the minimum separation on the initial relative inclination ({panel~B}), we see that the majority of systems are roughly consistent with the maximum eccentricity predicted by the torus-averaged approximation (cf. Eqs.~\eqref{eq:emax-sec}). They show a characteristic concentration of large eccentricities around initial inclinations $\cos i_0\approx0$, defining a narrow cone of inclinations within which large eccentricity excitations are possible in the torus-averaged approximation. As a reference, the minimum separation in the torus-averaged approximation is shown for initially circular binaries ($e_0=0$) at our smallest initial semi-major axis $a_0=10^2\,\rm AU$ (black line).  

\subsection{Strong oscillations}\label{sec:non-secular}
The torus-filling approximation described in the previous subsection fails to characterise the systems which undergo the strongest eccentricity oscillations and attain the smallest separations in our population. Instead, Fig.~\ref{fig:rmin-r0} shows that high eccentricities are achieved well beyond the narrow cone of initial inclinations predicted by the torus-filling approximation. 
Depending on the type of wide binary these high-eccentricity excitations can have important evolutionary implications. A few $0.1\,\%$ of the binaries in our population are driven to $r_{\rm min}<\rm R_\odot$ (red dashed line) where we would expect a stellar collision if the wide binary was composed of two solar-type main-sequence stars \citep[cf.,][]{Kaib2014}. In stellar binaries reaching $r_{\rm min}\gtrsim \rm R_\odot$ (red-shaded area) one or both members will fill their Roche-lobe at periapsis \citep{Eggleton1983} and transfer mass to their companion which may give rise to electromagnetic signals. About a few $0.01\,\%$ of the binaries are driven to separations which inevitably lead to a subsequent inspiral due to gravitational-wave emission (purple dashed line). If these binaries are composed of black holes or neutron stars their inspiral will lead to a merger that could result in a multi-band gravitational-wave detection (see below). Towards $a_0\sim10^5\,\rm AU$ we observe a depletion of large eccentricities because these loosely bound systems tend to be quickly disrupted by the pull of the Galactic tides (see Sec.~\ref{sec:disruptions}). 

Panel~C of Fig.~\ref{fig:rmin-r0} shows that the resulting fraction of extremely eccentric binaries leading to the aforementioned evolutionary outcomes are neither expected from the initial thermal distribution nor in the torus-filling approximation. Instead, we identify two reasons leading to a break-down of the latter and to extreme eccentricity excitations which are as follows.

Firstly, the maximum eccentricity in the torus-filling approximation is derived from a torus-averaged potential in which any short-term variation of $\Phi_{ij}$ is washed out. In reality, there are small but non-zero residuals $\delta\Phi_{ij}=\Phi_{ij}-\langle\Phi_{ij}\rangle$ along the trajectory of the binary barycentre which are neglected in the torus-averaged approximation. These residuals are peculiar near a high-eccentricity peak where the binaries only carry a tiny orbital angular momentum $L\propto\sqrt{1-e^2}$. Thus, even a small $\delta\Phi_{ij}$ could give rise to a torque which changes $L$ by the order of itself. For hierarchical triples in which the binary is perturbed by the tidal field of a distant tertiary companion it is well-known that these residuals can  result in a different maximum eccentricity than predicted from an orbit-averaged potential \citep{Antonini2014,Luo2016,Antonini2017,Grishin2018,Hamilton2023b}. For binaries moving in some arbitrary potential, we generalise previous findings and estimate the importance of this effect by considering the instantaneous torque exerted by the Galactic potential
\begin{align}
    \frac{d\mathbf{L}}{dt}&=-\mu\mathbf{r}\times\nabla\Phi|_{\mathbf{R}+\mathbf{r}}.
\end{align}
Introducing a set a of orthonormal basis vectors $\hat{\mathbf{e}}_r$, $\hat{\mathbf{e}}_L$, and $\hat{\mathbf{e}}_p$ that point along the relative separation of the binary, its orbital angular momentum, and a vector which is orthogonal to the latter two, respectively, the leading-order (quadrupole) contribution to the torque can be written as
\begin{align}
    \frac{d\mathbf{L}}{dt}&=\mu r^2\left(\Phi_{rp}\hat{\mathbf{e}}_L-\Phi_{rj}\hat{\mathbf{e}}_p\right),
\end{align}
where the potential derivatives $\Phi_{rp}$ and $\Phi_{rj}$ are evaluated at $\mathbf{R}$.
The torque is maximised at apoapsis $r=a(1+e)$ where in the limit $e\rightarrow1$ the timescale for changes in $L=\|\mathbf{L}\|$ evaluates to
\begin{equation}\label{eq:timescale-torque}
    \bigg\|\frac{1}{L}\frac{dL}{dt}\bigg\|^{-1}=\frac{\pi}{\sqrt{2}}\frac{\sqrt{1-e}}{T\|\Phi_{rp}\|}.
\end{equation}
If this timescale is shorter than the orbital period $T$ any secular treatment in which the potential is averaged over the orbital period or the ``outer" period inside the Galaxy ($1/\dot{\Omega}_\rho$, $1/\dot{\Omega}_\phi$, or $1/\dot{\Omega}_z$) formally breaks down \citep[c.f.,][]{HamiltonRafikovI,HamiltonRafikovII,BubPetrovich,HamiltonRafikovIII,Rasskazov2023} and must be replaced by a direct $N$-body integration like the one presented in this study. The condition for secular break-down can be rewritten in compact form as 
\begin{equation}\label{eq:secular-condition}
    \sqrt{1-e}\lesssim T^2\|\Phi_{rp}\|,
\end{equation}
where for some arbitrary orientation of the orbital frame $(\hat{\mathbf{e}}_r, \hat{\mathbf{e}}_p, \hat{\mathbf{e}}_L)$, $\|\Phi_{rp}\|$ can be as large as $\sim{\rm max}_{i,j=x,y,z}(\|\Phi_{ij}\|)$. Eq.~\eqref{eq:secular-condition} generalises the condition found by \cite{Antonini2014} for hierarchical triples to arbitrary perturbing potentials. One recovers the former by explicitly using $\Phi=-Gm_3/a_{\rm out}(1-e_{\rm out})$ where $m_3$ is the mass of the tertiary companion and $a_{\rm out}(1-e_{\rm out})$ is the periapsis of its orbit around the inner binary barycentre.

\begin{figure*}[h]
    \centering
    \includegraphics[width=.76\textwidth]{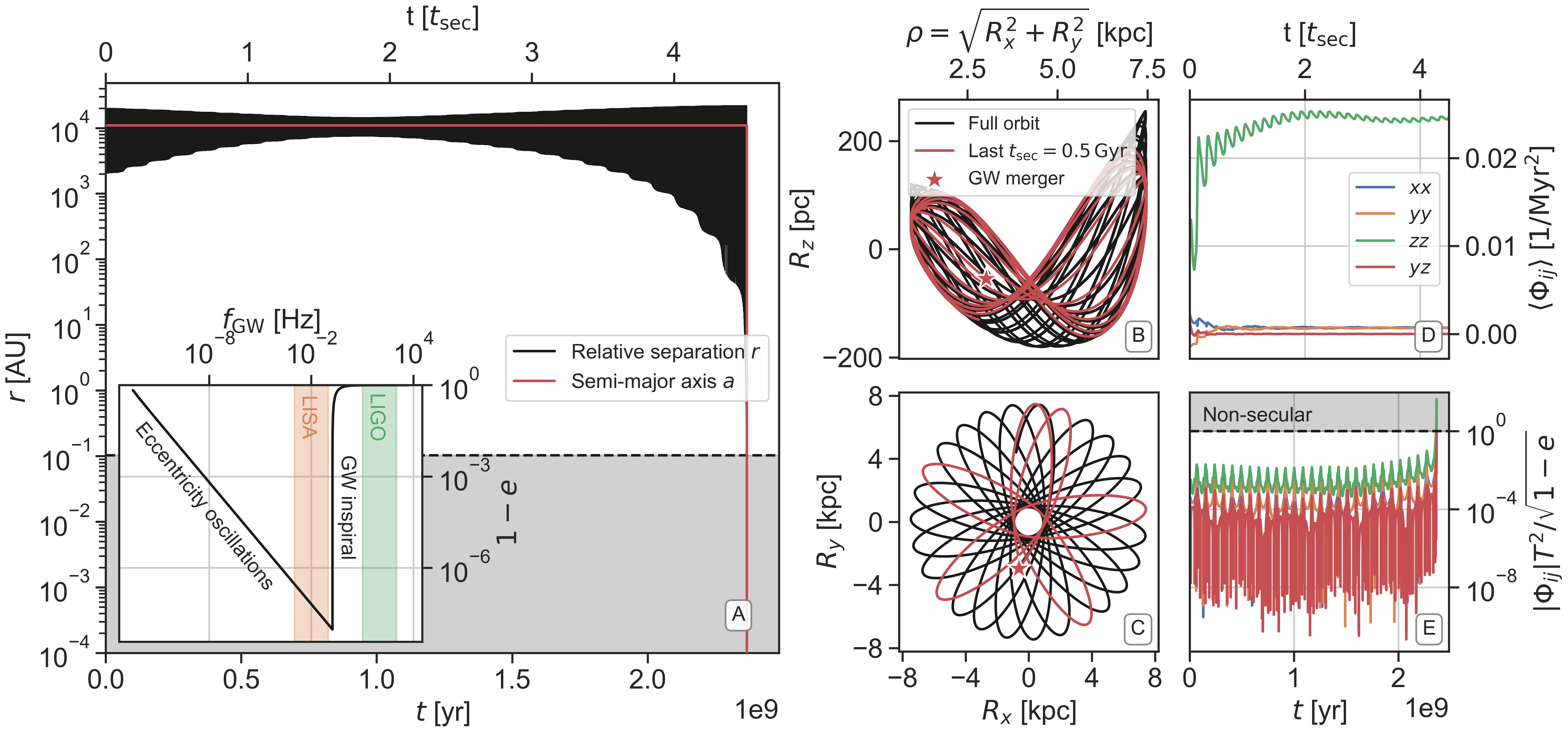}
    \caption{Same as Fig.~\ref{fig:secular-torus-filling} for a binary undergoing a non-secular high-eccentricity excitation after $t\approx2.3\,\rm Gyr$ followed by a gravitational-wave merger. The inset to {panel~A} displays the evolution of the binary eccentricity $e$ and the dominant harmonic of the gravitational-wave frequency $f_{\rm GW}$ \citep{Wen2003} through the bandwidths of LISA and LIGO/Virgo/Kagra.
    }
    \label{fig:non-secular-torus-filling}
\end{figure*}

In {panel~E} of Fig.~\ref{fig:secular-torus-filling} we can see that the previous example binary remains in the secular torus-filling regime as condition~\eqref{eq:secular-condition} remains unfulfilled and Eq.~\eqref{eq:emax-sec} accurately describes the maximum eccentricity. In contrast, we show in Fig.~\ref{fig:non-secular-torus-filling} a binary which is torus-filling in the sense of Sec.~\ref{sec:torus-secular} but whose evolution satisfies Eq.~\eqref{eq:secular-condition} and becomes non-secular near the high-eccentricity peak at $t\approx2.3\,\rm Gyr$. As a consequence, the tidal field torques the binary to an extreme eccentricity $1-e_{\rm max}\approx 10^{-8}$ where at periapsis gravitational wave radiation decouples the binary from the Galactic perturbation and causes an orbital decay and circularisation through the bandwidths of LISA and LIGO/Virgo/Kagra towards a merger (inset to {panel~A}).


We stress that substituting $\|\Phi_{rp}\|\rightarrow{\rm max}_{i,j=x,y,z}(\|\Phi_{ij}\|)$ in Eq.~\eqref{eq:secular-condition} must be seen as an order-of-magnitude calculation for the occurrence of high eccentricities due to secular break-down. In practice, the accuracy of the substitution for individual systems could be limited due to the facts that (i) the sign of $\Phi_{rp}$ may actually cause a torque to lower eccentricities, (ii) writing $\Phi_{rp}$ in terms of $\Phi_{i,j=x,y,z}$ with respect to the coordinates depends on the actual orientation of the binary, and (iii) Eq.~\eqref{eq:secular-condition} was derived from the instantaneous torque at apoapsis and does not guarantee that the binary maintains its (high) eccentricity value until the subsequent pericentre passage, but also (iv) the torus-filling approximation already fails if the timescale~\eqref{eq:timescale-torque} is less than the ``outer" period $\sim1/\dot{\Omega}_\phi$ inside the Galaxy (which is generally much longer than the binary period $T$). Nevertheless, we find that satisfying Eq.~\eqref{eq:secular-condition} to undergo secular break-down correlates well with the systems which experience extreme eccentricities that can no longer be described by the torus-filling approximation. 

\begin{figure*}[h]
    \centering
    \includegraphics[width=.76\textwidth]{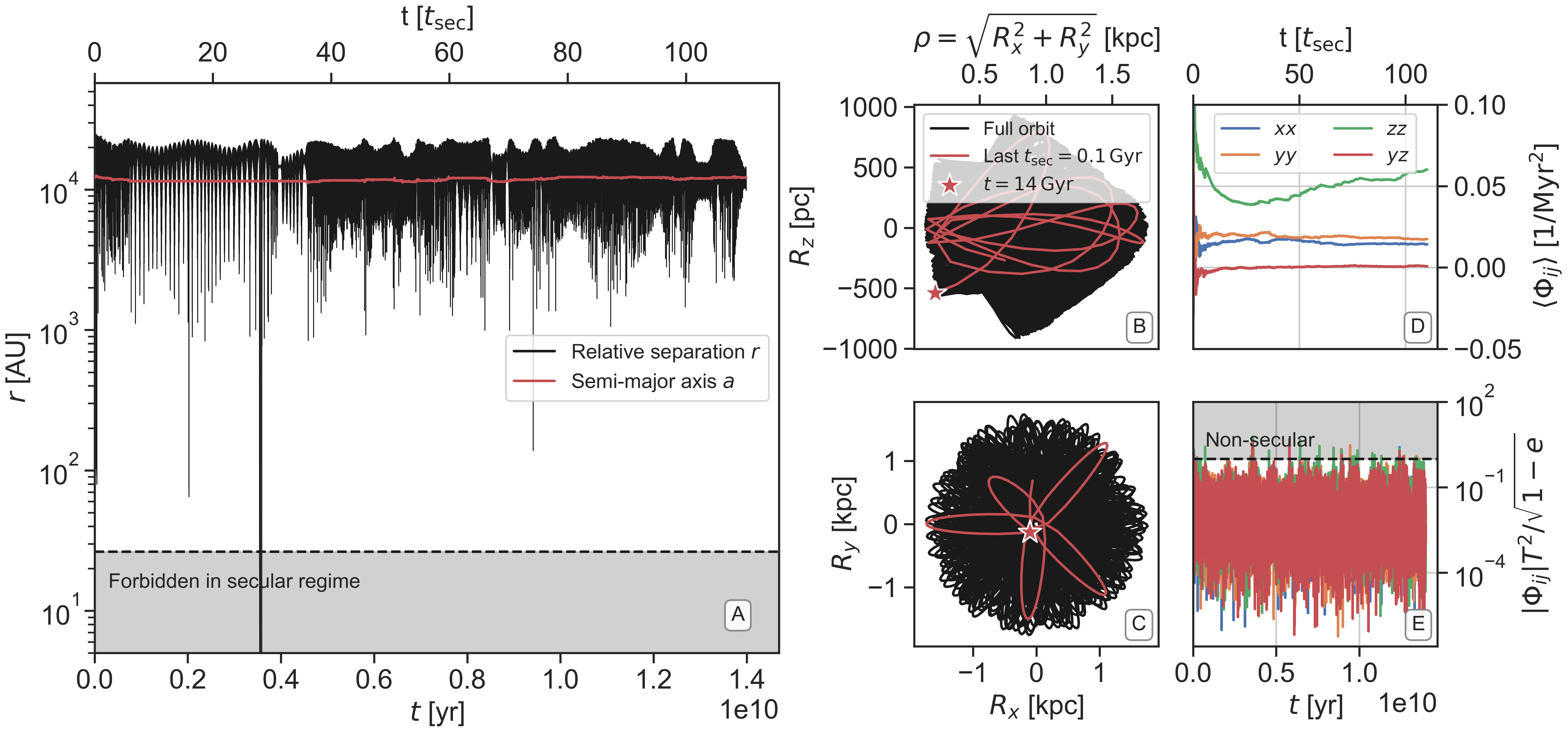}
    \caption{Same as Fig.~\ref{fig:secular-torus-filling} for a binary undergoing chaotic evolution.}
    \label{fig:non-secular-chaos}
\end{figure*}

The second reason for which the torus-filling approximation can break down is the emergence of chaos. The previous examples are characterised by $\dot{\Omega}_\phi t_{\rm sec}\gg1$ in which case within a secular timescale the tidal tensor components $\Phi_{ij}$ quickly converge to their torus-average, $\langle\Phi_{xx}\rangle=\langle\Phi_{yy}\rangle$, and $\langle\Phi_{ij}\rangle=0$ for $i\neq j$ ({panel~D} of Fig.~\ref{fig:secular-torus-filling} and~\ref{fig:non-secular-torus-filling}). As shown by \cite{HamiltonRafikovI,HamiltonRafikovII,HamiltonRafikovIII} this results in regular cycles of secular changes on a timescale $\mathcal{O}(t_{\rm sec})$. In contrast, the evolution can become chaotic if the azimuthal period of the barycentre becomes comparable to the secular timescale ($\dot{\Omega}_\phi t_{\rm sec}\approx1$) an example of which is shown in Fig.~\ref{fig:non-secular-chaos}. In that case, the averaged tidal tensor components fail to converge even on many secular timescales $t_{\rm sec}$ ({panel~D}) and the binary undergoes a random walk through phase space which eventually leads to extreme eccentricities. Equivalent findings were made for hierarchical triples and quadruples and binaries orbiting inside a cluster potential \citep{Petrovich2017,Hamers2017,Liu2019,BubPetrovich}.

\begin{figure*}[h]
    \centering
    \includegraphics[width=.76\textwidth]{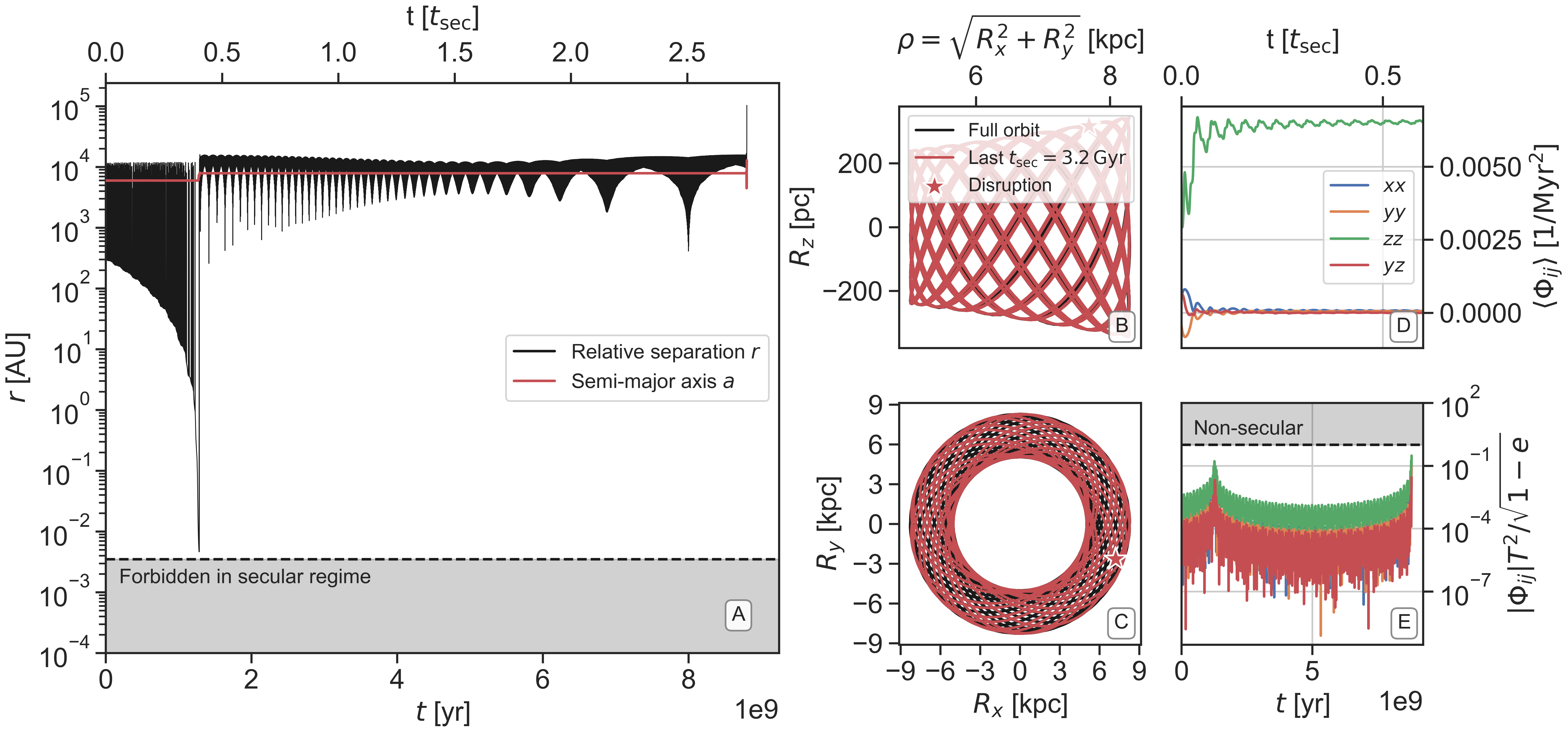}
    \caption{Same as Fig.~\ref{fig:secular-torus-filling} for a binary that is perturbed by stellar encounters and which gets disrupted after $t\approx9\,\rm Gyr$.}
    \label{fig:secular-encounters}
\end{figure*}

\subsection{Effect of stellar encounters}\label{sec:results-encounters}
Considering the effect of stellar encounters, panel~C of Fig.~\ref{fig:rmin-r0} shows that their inclusion marginally reduces the number of high-eccentricity excursions in our population. Comparing the simulations with and without stellar encounters, we find the numbers of stellar collisions ($r_{\rm min}<\rm R_\odot$) and gravitational-wave mergers get reduced by factors of $\sim2$ and $5$, respectively. It is noteworthy that we find almost no system which undergoes a stellar collision or gravitational-wave merger in both models. Instead, the actual evolutionary outcome is randomised. This is due to the fact that stellar encounters at each timestep add a small random perturbation to the relative velocity of the binary (see Sec.~\ref{sec:encounters}) which may either break a resonance that would have led to a high eccentricity due to Galactic tides alone or may actually be the cause that placed it on an resonant orbit. An example is shown in Fig.~\ref{fig:secular-encounters} where the binary undergoes large-eccentricity excitation on a resonant highly inclined trajectory ($\cos i_0\approx0$, see {panel~B} of Fig.~\ref{fig:rmin-r0}) until a stellar encounter at $t\approx1.5\,\rm Gyr$ breaks the resonance and quenches further perturbation. Finally, another encounter at $t=9.0\,\rm Gyr$ disrupts the binary. Disrupting binaries are studied in the following Sec.~\ref{sec:disruptions}. 

\subsection{Disruptions}\label{sec:disruptions}
\begin{figure}
    \centering
    \includegraphics[width=1\linewidth]{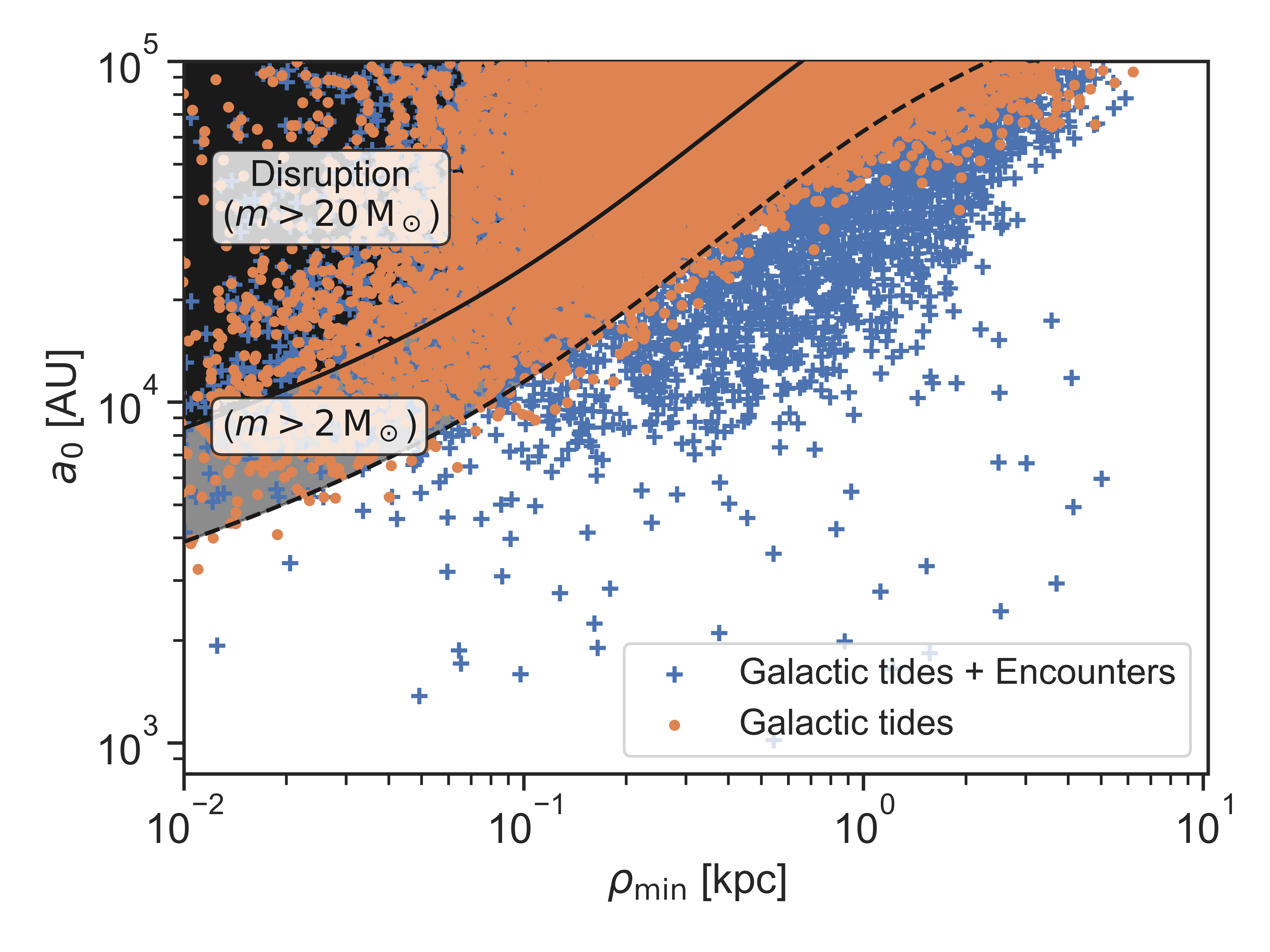}
    \caption{Initial semi-major axis $a_0$ and minimum Galactocentric radius $\rho_{\rm min}$ of wide binaries which get disrupted. Disruptions are defined when the binary separation at any time in the simulation exceeds $r=10a_0$ (see Sec.~\ref{sec:eom}). The minimum Galactocentric radius $\rho_{\rm min}$ is defined as the minimum of ${\rho}$ of the barycentre trajectory inside the Galaxy. Grey- and black-shaded regions correspond to Hill instability evaluated at that distance for binary masses $m>2\,\rm M_\odot$ and $m>20\,\rm M_\odot$, respectively. These limits correspond to the minimum and maximum mass in our population (see Sec.~\ref{sec:initial}).
    }
    \label{fig:disruptions}
\end{figure}

Owing to their low binding energy, wide binaries are susceptible to disruptions by Galactic tides and the (cumulative) effect of stellar encounters, which we find is the case for around $10\,\%$ of the binary population in either simulation. 
A useful quantity to understand whether a binary can get disrupted by the Galactic tides is set by the Jacobi radius $r_J$ given by \citep{BinneyTremaine,Yan-Fei2010}

\begin{equation}\label{eq:Jacobi1}
    r_J(\rho)=f(\rho)\left[\frac{m}{3M(\rho)}\right]^{1/3}\rho,
\end{equation}
where
\begin{equation}\label{eq:Jacobi2}
    f(\rho)=\left[1-\frac{1}{3}\frac{d\ln M(\rho)}{d\ln \rho}\right]^{1/3},
\end{equation}
and $M(\rho)$ is the mass of the Galaxy interior to $\rho$. The Jacobi radius given by Eqs.~\eqref{eq:Jacobi1} and~\eqref{eq:Jacobi2} defines the maximum extent of a satellite body orbiting on a circular trajectory inside a spherically symmetric host \citep{BinneyTremaine}. Our host Galaxy is neither spherically symmetric nor do we restrict our binary barycentres to circular trajectories. Nevertheless, we find that $r_J$ evaluated at the minimum Galactocentric radius $\rho_{\rm min}={\rm min}\left({\rho(t)}\right)$ along the barycentre trajectory demarcates well the region of stability of our simulated binaries. This is shown in Fig.~\ref{fig:disruptions} where the apoapsis of most disrupted binaries exceeds the Jacobi radius at their minimum Galactocentric radius ($r\approx2a_0>r_J$; solid/dashed lines). This result confirms previous findings of \cite{Yan-Fei2010} who found that $r\sim\mathcal{O}(r_J)$ sets a natural limit to the stability of wide binaries in the solar neighborhood.

Fig.~\ref{fig:disruptions} shows that exceptions to this stability criterion only occur if stellar encounters are taken into account. In that case, the velocity kicks imparted to the binary members can occasionally unbind them even if they are much closer to each other than the stability limit defined by the Jacobi radius.

\section{Rate calculations}\label{sec:rate}
Here, we estimate the galaxy-driven volumetric event rate of binary black hole (BBH) mergers, collisions of stars on the main-sequence (MSMS), and of binary white dwarfs (BWDs) in the local Universe. For this purpose, we adopt a simplistic formalism and do an order-of-magnitude estimate of the rate in the flavour of a Drake equation, which for BBH mergers reads
\begin{align}\label{eq:Drake}
\mathcal{R}_{\rm BBH} = \underbrace{\rho_{\rm MW}}_{\substack{\text{Number density} \\ \text{MW-like galaxies}}} &\times \underbrace{N_{\rm MW}}_{\substack{\text{Number of} \\ \text{stars in MW}}} \nonumber\\
\times\underbrace{A}_{\substack{\text{Massive pri-} \\ \text{mary fraction}}}  
&\times \underbrace{B}_{\substack{\text{Compan-} \\ \text{ion fraction}}} \nonumber\\
\times\underbrace{C}_{\substack{\text{Massive compan-} \\ \text{ion fraction}}} 
&\times \underbrace{D}_{\substack{\text{Wide BBH form-} \\ \text{ation fraction}}} \nonumber\\ 
\times\underbrace{E}_{\substack{\text{Galaxy-driven} \\ \text{merger fraction}}} 
&\times \underbrace{\frac{1}{14\,\rm Gyr}.}_{\substack{\text{per max. inte-} \\ \text{gration time}}}
\end{align}
We assume that the number density of Milky Way-like galaxies is $\rho_{\rm MW}=0.0116\,\rm Mpc^{-3}$ and that they host a total number of $N_{\rm MW}=1$~--~$4\times10^{11}$ stars \citep{Kopparapu2008}. Only a fraction $A$ of all stars is massive enough to form a black hole. Adopting a lower limit of $20\,\rm M_\odot$ and a \cite{Kroupa} initial mass function between $0.08$ and $150\,\rm M_\odot$ this fraction evaluates to $A=1.3\times10^{-3}$. The large majority $B=0.94$ of massive stars is accompanied by another star \citep{Moe2017}. In this calculation, we do not distinguish between companions that form an isolated binary, triple, or higher-order configuration and discuss this approximation in Sec.~\ref{sec:discussion}. Again, only a fraction $C$ of the companions is massive enough to form a black hole too. Because we are interested in the formation of wide BBHs, we can draw the mass ratio $q=m_2/m_1$ from the observed distribution $p(q)\propto q^{-2}$ between $0.1$ and $1.0$ for long-period binaries, which is nearly consistent with an independent sampling of the companion mass from the initial mass function \citep{Moe2017}. Thus, we find for the fraction in which both stars are massive ($m_{1,2}>20\,\rm M_\odot$) $A\times C=1.3\times10^{-4}$. To estimate the fraction $D$ of massive stellar binaries which successfully form wide BBHs in the simulated semi-major axis range $10^2$~--~$10^5\,\rm AU$, we run a population of massive binaries ($m_{1,2}>20\,\rm M_\odot$) from the zero-age-main-sequence with the rapid binary population synthesis code {\tt COMPAS} \citep{Riley2022}. Since wide BBHs originate from non-interacting stars their fraction is fully determined by the stellar wind model and the natal kick prescription. Here, we follow \cite{Vink2001} for the winds and the ``delayed" fallback supernova engine model of \cite{Fryer2012}. Assuming a log-uniform semi-major axis distribution between $10^{-1}$ and $10^5\,\rm AU$ \citep{Offner2023}, a thermal eccentricity distribution between 0 and 1 \citep{Moe2017}, a \cite{Kroupa} mass function for the primary star, and $p(q)\propto q^{-2}$ between $0.1$ and $1.0$ for the secondary, we find $D=0.41$ and $0.37$ at low ($Z=2\times10^{-4}$) and high metallicity ($Z=2\times10^{-2}$), respectively. Our simulations show that a fraction $E\approx5\times10^{-5}$ of these are driven to a gravitational-wave merger by the Galactic perturbation  (see panel~C of Fig.~\ref{fig:rmin-r0}). These assumptions result in a BBH merger rate $\mathcal{R}_{\rm BBH}\approx0.2$~--~$0.8\,\rm Gpc^{-3}yr^{-1}$ which amounts to a fraction $0.4$~--~$5\,\%$ of the observed rate $17.9$~--~$44\,\rm Gpc^{-3}yr^{-1}$ ($90\,\%$ C.L.) inferred from LIGO/Virgo/Kagra gravitational-wave detections \citep{Abbott2023}. 

To estimate the rate of collisions $\mathcal{R}_{\rm MSMS}$ of main-sequence stars we note that the wide binary fraction ($a\gtrsim10^2\,\rm AU$) of low-mass stars ($m_1=0.3$~--~$5\,\rm M_\odot$) has been observed to be around $10$ to $46\,\%$ which we can use to replace $B\times C\times D$ in Eq.~\eqref{eq:Drake} \citep{Offner2023}. For a \citep{Kroupa} mass function, the fraction of primaries in the range $m_1=0.3$~--~$5\,\rm M_\odot$ evaluates to $A=0.38$, and our simulation shows that a fraction $E\approx2\times10^{-3}$ of binaries fall below $r_{\rm min}<1\,\rm R_\odot$. Thus, we infer a total rate $\mathcal{R}_{\rm MSMS}\approx0.6$~--~$11.6\times10^{-5}\,\rm Mpc^{-3}yr^{-1}$ which is consistent with recent measurements of the rate $7.8^{+6.5}_{-3.7}\times10^{-5}\,\rm Mpc^{-3}yr^{-1}$ of Luminous Red novae \citep[][]{Karambelkar2023} and agrees with previous estimates from \cite{Kaib2014}.

Lastly, for the rate of binary white dwarf collisions we simply reduce the previous rate estimate for MS collisions using $E\approx5\times10^{-5}$ to account for the lower probability to get $r_{\rm min}<10^{-2}\,\rm R_\odot$. In order to make a meaningful comparison to observations we additionally impose a minimum mass of the primary of $0.8\,\rm M_\odot$ for the collision to produce enough radioactive nickel to be visible as a type Ia SN, as also assumed by \cite{Ruiter2011}. For the \citet{Kroupa} mass function this introduces another reduction by a factor of 50, so that $\mathcal{R}_{\rm BWD}$ amounts to a fraction $0.01$~--~$0.1\,\%$ of the measured rate $4.1\times10^{-5}\,\rm Mpc^{-3}yr^{-1}$ of Type Ia SNe \citep{Toy2023}.

We stress that Eq.~\eqref{eq:Drake} leads to an order-of-magnitude rate estimate that takes into account neither how different other potentials than the Milky Way contribute (see Sec.~\ref{sec:discussion}), nor how the merger fraction depends on the mass of the binary members. Furthermore, it does not properly convolve the star-formation rate as a function of time with the delay time distribution for a particular event. A more accurate rate estimate, which specifically depends on the type of event, is beyond the scope of this paper and should be addressed in the future. However, we have checked that the merger fraction is nearly independent of the input masses of the binaries and that the typical delay time of the events, which is a few $\rm Gyr$, coincides with the lookback time of the peak in star formation \citep{Madau2014}. 




\section{Conclusions}\label{sec:discussion}
In this study, we have shown that the torque exerted by the gravitational potential of the Milky Way can induce eccentricity oscillations of wide binaries orbiting inside the Galactic field. Largely relying on one or more of the assumptions \ref{item:1}~--~\ref{item:3} in Sec.~\ref{sec:Intro}, this has been known for decades \citep[e.g.,][]{Heisler1986} but we find that the maximum possible eccentricity excitation has been severely underestimated. By employing a direct $N$-body integrator which relies on none of the previous assumptions we have shown that near-unity eccentricities ($1-e\lesssim10^{-8}$) are possible in two different dynamical regimes. In the weak regime (Sec.~\ref{sec:torus-secular}) assumptions \ref{item:1}~--~\ref{item:2} are well-satisfied and large eccentricities are only possible within a narrow cone $\cos i_0\approx0$ of initial inclinations between the binary plane and Galactic plane (Fig.~\ref{fig:rmin-r0}), similar to the von Zeipel-Kozai-Lidov effect in hierarchical triples \citep{Naoz2016} and binaries inside stellar clusters \citep{HamiltonRafikovI,HamiltonRafikovII,BubPetrovich,HamiltonRafikovIII}. This weak regime breaks down if the Galactic torque acts on a timescale less than the orbital period (Eq.~\eqref{eq:secular-condition}) or if chaos emerges when the binary orbits inside the Galaxy on a timescale similar to the secular timescale (Sec.~\ref{sec:non-secular}). In either case, extreme eccentricities can be achieved beyond the cone of inclinations derived from the secular approximation. Depending on the type of wide binary, this dynamics may have important implications for the occurence of stellar collisions, of white dwarf collisions, and binary black hole mergers and may contribute significantly to the observed rate of these events (Sec.~\ref{sec:rate}).

We conclude this study by highlighting several aspects that need to be addressed in future work. 

\begin{enumerate}
    \item We have focused our investigation on binaries orbiting inside a Milky Way potential which was approximated as being axisymmetric and stationary (Sec.~\ref{sec:eom}). It is not obvious whether including the non-axisymmetric and time-dependent density field of the Milky Way (including its bar and spiral structure) will change our findings. As with stellar encounters, it could either tend to break the resonance or introduce additional perturbations $\delta\Phi_{ij}$ of the tidal field (see Sec.~\ref{sec:non-secular}) which may actually drive the binaries into the non-secular regime. Regarding other galactic potentials we do not have any reason to expect that the dynamical evolution would be qualitatively different. However, the potential of, e.g., elliptical, irregular, and dwarf galaxies, may significantly alter the fraction of each events quantitatively due to potentially different initial binary populations, the strength of Galactic torques, and the abundance of chaotic orbits. In addition, the time-evolution of such potentials could affect our results in a non-trivial way. In our analysis we find a relatively large span of delay times between the start of the simulation and the occurrence of a particular event. For instance, the gravitational-wave mergers cover a total range of delays between $\sim0.1$ and $14.1\,\rm Gyr$. The bulk of systems merge after a few $\rm Gyr$, with a median value ($75\,\%$~--~quantile) of $5.1\,\rm Gyr$ ($10.5\,\rm Gyr$), which, if contributing to the present-day local merger rate, would correspond to a birth redshift of $z\approx0.5$ ($2.0$). Thus, a complete Galactic model would need to take into account the evolution over at least $0\lesssim z\lesssim2$.
    \item When low-mass stars approach each other closely at periapsis by the mechanism explored in this study, they interact with each other through stellar tides \citep[e.g.,][]{Zahn1975,Hut1981}. As shown by \citet{Kaib2014} the dissipated energy by tides can convert a wide binary at small periapsis to a close or contact binary which is shielded against further perturbation from the Galaxy. Exploring a set of various tidal models for the (uncertain) amount of dissipated energy, \citet{Kaib2014} found that the formation of close binaries reduces the number of Galactically-driven stellar collisions by a factor of order unity, compared to no-tide evolution, for their fiducial dissipation models of polytropic stars \citep{Lee1986,Press1977} and a factor of around ten for some artificially extreme models. Thus, for any plausible tidal model, they find enough stellar mergers in their simulations to explain the observed rate of stellar collisions in the Milky Way. For the purpose of focusing on the point mass dynamics we have ignored the effect of tides in our work. Yet, applying the reduction factors found by \citet{Kaib2014} our rate of stellar collisions (Sec.~\ref{sec:rate}) is lowered by a factor of a few at most.  
    \item Unlike low-mass stars, their massive counterparts ($m_{1,2}\gtrsim8\,\rm M_\odot$) evolve within a few Myr to neutron stars or black holes which is much shorter than the secular timescale $t_{\rm sec}$ at which the Galaxy could drive the binaries to a close periapsis. Therefore, it is justified to neglect the effect of stellar tides for wide massive binaries. However, wide orbits of massive stars are found most often to be the outer orbit of a hierarchical triple or higher-order configuration \citep{Moe2017}, i.e., one or both wide binary members exhibit additional sub-companions. In addition, compact objects may experience a significant natal kick when they form from massive stars \citep{Tauris2023} which risks disrupting the loosely bound wide orbit. The effects of massive star multiplicity and natal kicks introduce uncertainties to our rate estimates of binary black hole mergers (Sec.~\ref{sec:rate}). In general, natal kicks are caused by asymmetric mass ejection and neutrino emission during the core-collapse supernova explosions. 
    It is thought that black holes tend to accrete all the stellar material so that their kicks are mainly determined by the asymmetric neutrino emission leading to typical kicks of several $\rm km\,s^{-1}$ at most \citep{Janka2024,Burrows2024}, which is supported by recent observations of orbits hosting a black hole \citep{Shenar2022,Vigna-Gomez2023,Burdge2024}. Despite the kick uncertainties due to the exact explosion mechanisms and the amount of matter accreted onto the black hole \citep{Mandel2016}, it is hence reasonable to assume that for many black holes the kick velocity is smaller than the typical orbital velocity of wide binaries (Eq.~\eqref{eq:orbital-velocity}) so that a significant number of wide massive binary stars successfully form wide binary black holes \citep{Olejak2020}. 
    To estimate the impact of natal kicks on the formation of wide BBHs, we initiate a population of $N_0=10^5$ massive binary stars ($m_1=m_2=10\,\rm M_\odot$) with a log-uniform semi-major axis distribution between $10^{-1}$ and $10^5\,\rm AU$, a thermal eccentricity distribution, and isotropic orbital angles. We consecutively apply isotropic natal kicks at random orbital phases to each of the stars and update the orbital elements accordingly. The magnitude of the natal kicks is drawn from a Maxwellian velocity distribution with dispersion $\sigma_{\rm kick}=0.1$, $0.5$, $1.0$, $5.0$, and $10.0\,\rm km\,s^{-1}$, respectively. Figure~\ref{fig:BBH-distribution} shows the final semi-major axis distribution of surviving BBHs indicating that for $\sigma_{\rm kick}\lesssim\mathcal{O}(1)\,\rm km\,s^{-1}$ a significant fraction can be expected throughout the wide binary range explored in this work. Conversely, three-dimensional supernova simulations consistently predict large natal kicks for neutron stars (typically a few hundred $\rm km\,s^{-1}$) which are mainly driven by the asymmetric mass ejection \citep{Janka2024} and substantiated by the observation of high pulsar velocities \citep[e.g.,][]{Hobbs2005}. This is much larger than the typical orbital velocity of wide binaries which makes the successful formation of bound wide binary neutron stars or black hole-neutron star binaries extremely unlikely. Additionally, wide compact object binaries might form from initially unbound pairs of single objects due to chance encounters during the dissolution phase of young star clusters \citep{Kouwenhoven2010}.

    \begin{figure}
    \centering
    \includegraphics[width=1\linewidth]{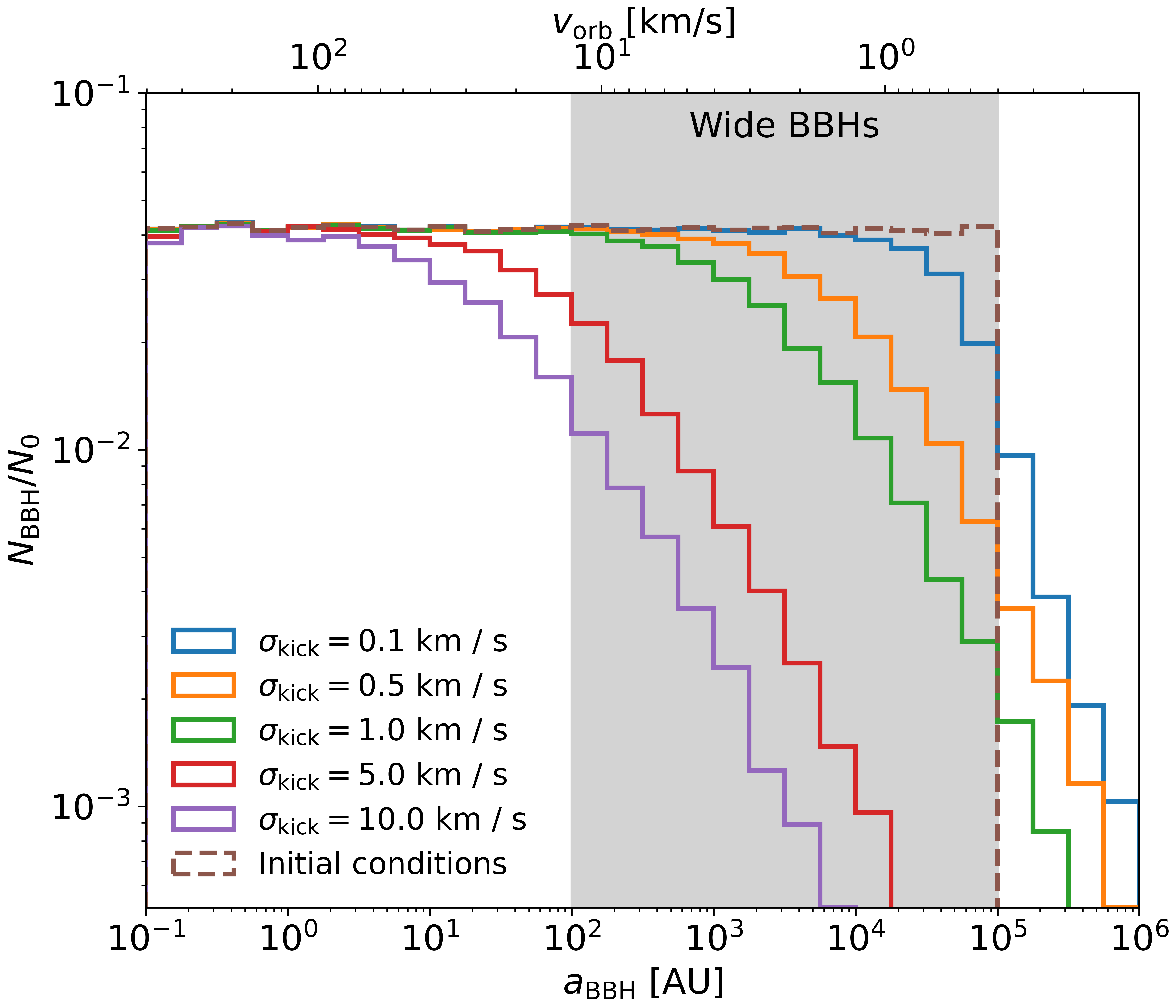}
    \caption{The effect of natal kicks on the formation of wide BBHs. The dashed line shows that the initial distribution of $N_0=10^5$ wide massive binaries ($m_1=m_2=10\,\rm M_\odot$) which consecutively receive two natal kicks at BH formation whose magnitudes are drawn from a Maxwellian velocity distribution with dispersion $\sigma_{\rm kick}$. Solid lines show the semi-major axis distribution of all surviving BBHs and the grey-shaded area highlights the range of values simulated in this work. As a reference, we also show the orbital velocity of the BBHs (Eq.~\eqref{eq:orbital-velocity}). 
    }
    \label{fig:BBH-distribution}
    \end{figure}

    \item Regarding multiplicity, while it is plausible that about $20\,\%$ of massive stellar triples and higher-order configurations could become wide binary black holes after the sub-companions are lost because of mergers during stellar evolution \citep{Sana2012,Stegmann2022}, a non-negligible number of wide binary black holes may form while retaining their sub-companions. \citet{Evgeni2022} have shown that a wide tertiary companion which is gently driven towards the inner binary through Galactic tides can trigger von Zeipel-Kozai-Lidov oscillations of the latter. For wide tertiary companions which are driven to a near-radial orbit through the mechanism studied in our work, we may additionally expect systems that undergo chaotic evolution at close periapsis passage similar to binary-single star scattering experiments \citep[e.g.,][]{Samsing2014} which likely lead to the ejection of one of the components or the merger of two. Thus, we anticipate that the inclusion of additional tertiary or higher-order companions introduces further complexity to the problem, potentially enlarging the parameter space for mergers. 

    We emphasise that the proposed channel to form compact object mergers from wide binaries (or higher-order configurations with a wide outer orbit) is appealing from a theoretical point of view. Since the wide binary stars evolve far apart from each other, they are not subject to the systematic uncertainties associated with the stellar evolution of interacting short-period massive binary stars that have been proposed to be the origin of compact object mergers; either through common-envelope evolution \citep{Postnov2014,Belczynski2016,Eldridge2016,Lipunov2017}, a stable mass transfer episode \citep{vandenHeuvel2017,Inayoshi2017,Neijssel2019,Bavera2021,Marchant2021,Gallegos-Garcia2021,Olejak2021,vanSon2022}, or chemically homogeneous evolution \citep{deMink2016,MandeldeMink2016,Marchant2016,duBuisson2020,Riley2021}. Instead, the uncertainty reduces to the initial properties and the formation of wide binaries \citep{El-Badry2024}, the orbital widening due to mass-loss by stellar winds if the stars are metal-rich \citep[][]{Vink2001,Bjorklund2021}, and to the remnant mass function and natal kick prescription of (effectively single) massive stars \citep{Heger2003}. 

    \item Finally, we point out that it would be worthwhile to use our integrator to investigate the evolution of the eccentricity distribution of wide binaries in the solar neighbourhood. \cite{Tokovinin2020} and \cite{Hwang2022} found that the wide binaries astrometically identified in the \textit{Gaia} DR2 sample show evidence for a super-thermal distribution, i.e., $p(e)\propto e^\alpha$ where $\alpha>1$. Working with the secular approximation, \cite{Modak2023} and \cite{Hamilton2023} have shown that neither Galactic tides nor stellar encounters could transform a thermal ($\alpha=1$) or sub-thermal ($\alpha<1$) to the observed distribution, concluding that dynamical effects cannot drive the binaries away from thermal equilibrium, hence they must already form with a super-thermal distribution. Whether their results withstand a direct $N$-body integration can be tested with our formalism. 
\end{enumerate}

\section*{Acknowledgments}
We thank Fabio Antonini, Stephen Justham, Re'em Sari, Hans-Thomas Janka, Kareem El-Badry, Chris Hamilton, Rainer Spurzem, Aleksandra Olejak, and Evgeni Grishin for useful input and discussions. TW acknowledges support from NASA ATP grant 80NSSC24K0768. L.Z. acknowledges support from  ERC Starting Grant No. 121817–BlackHoleMergs.

\newpage

\section*{Software and third party data repository citations} \label{sec:cite}
\facilities{Max Planck Computing and Data Facility (MPCDF)}
\software{{\tt MSTAR} \citep{Rantala2020,Mannerkoski2023},  
          {\tt gala} \citep{gala}, 
          {\tt cogsworth} (Wagg et al.~in prep.),
          {\tt COMPAS} \citep{Riley2022}
          }

\bibliography{sample631}{}
\bibliographystyle{aasjournal}



\end{document}